\documentclass[
  notitlepage,
  twocolumn,
  preprintnumbers,
  longbibliography,
  superscriptaddress,
  aps,
  prb,
  amsmath,
  amssymb,
]{revtex4-2}
\usepackage{amsmath, amssymb, amsthm}

\newtheorem{lemma}{Lemma}

\usepackage{subcaption}
\usepackage{listings}
\usepackage{xcolor}
\definecolor{codebg}{HTML}{F7F7F7}   
\definecolor{codeframe}{HTML}{DDDDDD}
\definecolor{pykw}{HTML}{0000AA}     
\definecolor{pybuilt}{HTML}{008000}  
\definecolor{pystring}{HTML}{BA2121} 
\definecolor{pydoc}{HTML}{408040}    
\definecolor{pylinen}{HTML}{AAAAAA}  

\lstdefinestyle{mintedpy}{
  language=Python,
  backgroundcolor=\color{codebg},
  frame=single,
  rulecolor=\color{codeframe},
  framerule=0.4pt,
  basicstyle=\ttfamily\small,
  keywordstyle=\color{pykw}\bfseries,
  commentstyle=\color{pydoc}\itshape,
  stringstyle=\color{pystring},
  showstringspaces=false,
  tabsize=4,
  breaklines=true,
  numbers=left,
  numberstyle=\tiny\color{pylinen},
  numbersep=8pt,
  mathescape=true,
  aboveskip=0.7\baselineskip,
  belowskip=0.7\baselineskip,
}

\lstdefinestyle{mintedpy-doc}{
  style=mintedpy,
  moredelim=**[is][\color{pydoc}\itshape]{"""}{"""},
}

\usepackage{graphicx}
\usepackage{multirow}
\usepackage{CJKutf8}
\usepackage[justification=raggedright,singlelinecheck=false]{caption}
\usepackage{subcaption}
\usepackage[margin=1in]{geometry} 

\usepackage{booktabs} 
\usepackage{physics}
\usepackage{ifthen}
\usepackage{braket}
\usepackage[normalem]{ulem}

\usepackage{amsmath}
\usepackage{amsthm, amssymb}
\usepackage{slashed} 
\usepackage{graphicx}
\usepackage{xcolor}
\usepackage{bm}
\usepackage[
  pdfpagelabels,
  plainpages=false,
  bookmarks=true,
  colorlinks=true,
  linkcolor=red,
  urlcolor=blue,
  citecolor=blue
]{hyperref}
\usepackage{dsfont}
\usepackage{changepage}
\usepackage{array}
\usepackage{bbm}
\usepackage[cal=boondox]{mathalpha}

\renewcommand{\vec}[1]{\bm{\mathbf{#1}}}
\newcommand{\Z}{\mathbb{Z}}

\newcommand{\loc}{\text{loc}}

\usepackage{xcolor}

\renewcommand{\O}{\mathcal{O}}
\newcommand{\h}{\mathcal{h}}
\begin{document}
\title{Real-Time Dynamics in Two Dimensions with Tensor Network States via Time-Dependent Variational Monte Carlo}

\author{Yantao Wu}
\email{yantaow@iphy.ac.cn}
\affiliation{Institute of Physics, Chinese Academy of Sciences, Beijing 100190, China}
\author{Jannes Nys}
\affiliation{Institute for Theoretical Physics, ETH Zürich, 8093 Zürich, Switzerland}

\date{\today}
\begin{abstract}
Reliably simulating two-dimensional many-body quantum dynamics with projected entangled pair states (PEPS) has long been a difficult challenge. 
In this work, we overcome this barrier for low-energy quantum dynamics by developing a stable and efficient time-dependent variational Monte Carlo (tVMC) framework for PEPS. 
By analytically removing all gauge redundancies of the PEPS manifold and exploiting tensor locality, we obtain a numerically well-conditioned tVMC equation, yielding robust solutions using the efficient Cholesky decomposition.
This enables long-time evolution in previously inaccessible regimes.
We explain how the difficulties in the traditional approach, particularly those associated with gauge redundancies, are resolved within tVMC.
We demonstrate the power and generality of the method through five representative real-time local quench dynamics in two dimensions: 
(I) chiral edge propagation in a free-fermion Chern insulator; 
(II) vison propagation in a pure $\Z_2$ gauge theory;
(III) vison confinement dynamics in a $\mathbb{Z}_2$ lattice gauge theory coupled to Higgs field;
(IV) fractionalized charge transport in a fractional Chern insulator; 
and (V) superfluidity and critical velocity in interacting bosons. 
All simulations are performed on $\ge 10\times 10$ lattices with evolution times beyond $T = 10$ using modest computational resources.
Where exact benchmarks exist (case I), PEPS–tVMC matches free-fermion dynamics with high accuracy up to $T = 12$.
In addition, we also simulate the paradigmatic dynamics of the Ising model following a global quench at the critical transverse field, and obtain, with modest bond dimension, agreement with previous results.
These results demonstrate PEPS–tVMC as a practical and versatile tool for real-time quantum dynamics in two dimensions. 
The method significantly extends the reach of classical tensor-network simulations for studying elementary excitations in quantum many-body systems in real-time and provides a valuable computational counterpart to emerging quantum simulators.
As a by-product in the development, we also present a new form of minSR, which is more stable and offers a new perspective on tVMC.
\end{abstract}
\maketitle

\section{Introduction}
Understanding real-time dynamics of interacting quantum many-body systems in two dimensions (2D) is a fundamental challenge that spans condensed-matter physics, particle physics, and quantum simulation. 
While powerful classical methods exist for ground states—including tensor networks \cite{cirac2021matrix, xiang2023tensor} and quantum Monte Carlo \cite{Sandvik2010QMC, BeccaSorella2017QMCBook}, real-time evolution in 2D remains notoriously difficult. 
Exact diagonalization is limited to small clusters; quantum Monte Carlo suffers from a dynamical sign problem.
Recently, the combination of time-dependent variational Monte Carlo (tVMC)~\cite{tvmc1, tvmc2, tvmc3, tvmc_4}, a continuous- and real-time generalization of the stochastic reconfiguration (SR) method \cite{SR1998, sorella2001generalized,sorella2017}, with neural quantum states (NQS) \cite{nqs} has emerged as a promising alternative~\cite{nqstvmc_3,nqstvmc_4,nqstvmc_7,nqstvmc_8}.
Here, we refer to tVMC as a Monte Carlo formulation of the time-dependent variational principle (TDVP)~\cite{tvmc1, tvmc2}.
NQS provides remarkable variational flexibility~\cite{sharir2022neural, hu2025efficient, sinibaldi2025non}, while tVMC offers a principled formulation of real-time evolution.
The combination has also motivated alternative approaches to dynamics within the VMC framework~\cite{nqstvmc_4, nqstvmc_6, nqstvmc_8, nqstvmc_9}.
At present, extending NQS–tVMC to large 2D systems remains an active area of research, where one must balance expressive parametrizations with efficient and stable optimization~\cite{Lange_2024}.

Projected entangled pair states (PEPS) \cite{PEPS2004} provide a complementary variational family whose locality and multi-linearity make them particularly appealing for tVMC.
Within VMC, only a single layer of PEPS tensors needs to be contracted, making the method substantially more efficient than conventional PEPS update schemes~\cite{full_update_1, full_update_2, algorithm_finite_peps}.
Furthermore, VMC allows parallel accumulation of samples, making the approach well suited for modern GPU architectures.
In fact, PEPS-VMC has recently achieved considerable success for ground states, giving competitive results for the ground states of frustrated spins \cite{liu2022gapless, liu2022emergence, liu2024quantum}, the Fermi-Hubbard model \cite{liu2025hubbard}, and lattice gauge theory \cite{wu2025accurate}, owing to a foundational breakthrough in the sampling method of PEPS \cite{liu2021}.

Although PEPS obey an area law for entanglement entropy~\cite{eisert2010area} and cannot represent volume-law states produced by fully thermalizing dynamics, they are well suited for \emph{local quenches} in which the injected energy does not scale with system size.
In this latter regime, the energy density remains zero (in the thermodynamic limit) and we expect the dynamics to preserve area-law entanglement~\cite{calabrese2016quench, wu2020dissipative}.
Recent theoretical studies also suggest area-law entanglement during local quenches \cite{garratt2026quantum,grover2026hierarchical}.
Thus, PEPS naturally targets low-energy local-quench dynamics --- precisely the regime relevant for elementary excitations and real-time response functions, and directly related to experiments including scanning tunneling microscopy \cite{Fischer2007STMReview}, angle-resolved photoemission spectroscopy \cite{Sobota2021ARPESReview}, particle scattering processes, and more.
A controlled method capable of simulating such dynamics provides a valuable tool for studying 2D quantum systems.

While dynamical simulations with matrix product states (MPS) are enormously successful in 1D \cite{TDVP, TEBD, WII, iTEBD}, the situation with PEPS in 2D is much more challenging. 
In contrast to the extensive 1D literature, only a small number of works—represented by Refs. \cite{czarnik2018time, czarnik2019time,hubig2019time,hubig2020evaluation,lin2021efficient,dziarmaga2021time,dziarmaga2022time,ponnaganti2022real,tindall2023gauging,espinoza2024spectral,motoyama2025tenes,dai2025fermionic}, have reported real-time simulations using PEPS or infinite PEPS (iPEPS), illustrating the general difficulty of the problem.
Traditional approaches to dynamical PEPS have relied on full-update-like methods based on double-layer contraction \cite{full_update_1, full_update_2, algorithm_finite_peps}, which face several challenges: (a) high cost of double-layer contraction, (b) difficulty in fixing the PEPS gauge freedom, and (c) loss of positivity in the quantum metric approximately obtained from the contraction of the double-layer environment.
In Sec. \ref{subsec:comparison}, we explain that, within PEPS--tVMC, the last two issues are naturally resolved while the first one is significantly alleviated, at the expense of manageable statistical noise.
We also note that, PEPS time-evolution, based on simple updates (SU) \cite{simple_update}, has recently experienced an exciting renaissance due to a new interpretation and improvement via belief propagation \cite{alkabetz2021belief,guo2023blockbp,evenbly2026loop}, giving impressive results on complex dynamics problems in tandem with recent experiments \cite{tindall2024efficient,begusic2024fast,tindall2025dynamics,lee2025scalable,luchnikov2024largescale}. 
We expect such progress to continue, co-growing with the development of quantum computers.

Despite the long development of PEPS \cite{cirac2021matrix}, their combination with tVMC for real-time dynamics has received virtually no attention, aside from our recent exploratory attempt in a simple setting \cite{wu2025accurate}.
Here, we develop PEPS--tVMC systematically, analyze its structure, and demonstrate its ability to simulate large-scale 2D dynamics.
Our approach builds on time-dependent Variational Monte Carlo ~\cite{tvmc1, tvmc2}.
We identify two ways to remove gauge redundancies in PEPS dynamics: one in the standard TDVP form, and another one in its dual minSR form \cite{minsr}.
After gauge removal, the TDVP equation exhibits remarkable numerical stability: the TDVP residual remains as low as $10^{-9}$ to $10^{-25}$ in the examples studied. 
As a result, PEPS—long viewed as difficult for real-time evolution—can in fact provide a powerful and controllable variational framework for 2D low-energy dynamics.
This resolves long-standing challenges that have hindered dynamical simulations with PEPS since their introduction \cite{PEPS2004}. 
We illustrate the versatility of PEPS--tVMC through six physically motivated examples, which illustrate that the method is not specialized to a single model class.

The paper is organized as follows.
In Sec.~\ref{sec:tvmc}, we review the tVMC framework and set up the notation.
In Sec.~\ref{sec:sr_peps}, we analyze the structure of PEPS--tVMC and present the ingredients necessary for its stability and efficiency.
In Sec.~\ref{sec:results}, we demonstrate the method through six representative examples.
In Sec.~\ref{sec:discussion} and \ref{sec:open}, we summarize our findings and outline future directions.
\section{Time-Dependent Variational Monte Carlo}
\label{sec:tvmc}

Consider a many-body lattice quantum system with computational basis
$\vec s = (s_1, \dots, s_N)$ with local Hilbert space dimension $d$.
To perform a variational simulation, we adopt a holomorphic parametrization of the \textit{unnormalized} quantum state
$\Psi_{\vec\theta}(\vec s)$ with $N_p$ complex variational parameters
$\vec\theta = \{\theta_\alpha : \alpha = 1,2,\dots,N_p\}$.
At time $t$, the exact evolution under a Hamiltonian $H$ for a short interval $dt$
induces an evolution from $\vec\theta(t)$ to $\vec\theta(t+dt)$ that minimizes the Fubini--Study distance~\cite{QGT} between
$\Psi_{\vec\theta(t+dt)}$ and $e^{-iH dt}\Psi_{\vec\theta(t)}$.
For infinitesimal $dt$, this yields the TDVP equations \cite{tvmc1, TDVP}:
\begin{equation}
  \sum_{\beta=1}^{N_p} S_{\alpha\beta}\,\dot\theta_\beta = -i\, g_\alpha ,
  \label{eq:sr}
\end{equation}
where the gram matrix or quantum geometric tensor (QGT) $S_{\alpha\beta}$ and the energy gradient $g_\alpha$ are given by (up to normalization factors)
\begin{align}
    S_{\alpha\beta} &= \braket{\partial_\alpha \Psi|\partial_\beta \Psi} - \braket{\partial_\alpha \Psi | \Psi} \braket{\Psi | \partial_\beta \Psi} ,  \\
    g_\alpha &= \braket{\partial_\alpha \Psi | H | \Psi} -  \braket{\partial_\alpha \Psi | \Psi} \braket{\Psi | H |\Psi}.
\end{align}
These quantities can be written in terms of expectation values over the Born distribution $P_{\vec\theta}(\vec s) = |\Psi_{\vec\theta}(\vec s)|^2/\sum_{\vec s}|\Psi_{\vec\theta}(\vec s)|^2$
\begin{align}
  S_{\alpha\beta}
  &= [O^*_\alpha(\vec s) O_\beta(\vec s)]
     - [O^*_\alpha(\vec s)] [O_\beta(\vec s)],
  \label{eq:S}
  \\
  g_\alpha
  &= [O^*_\alpha(\vec s)\,H_\loc(\vec s)]
     - [O^*_\alpha(\vec s)] [H_\loc(\vec s)].
  \label{eq:g}
\end{align}
Here $[\cdot]$ denotes a statistical average over
$P_{\vec\theta}(\vec s)$, 
$H_\loc(\vec s) = \braket{\vec s|H|\Psi}/\braket{\vec s|\Psi}$ is the local energy,
and
\begin{equation}
O_\alpha(\vec s) \equiv
\frac{\partial\log\Psi_{\vec\theta}(\vec s)}
     {\partial\theta_\alpha}
\end{equation}
is the log-derivative.

In practice, $S$ and $\vec g$ are evaluated stochastically by sampling configurations
$\{\vec s\}$ from $P_{\vec\theta}$.
Given $N_s$ samples, we form an $N_s\times N_p$ matrix $O$ with elements
$O_{s\alpha} = O_\alpha(\vec s)$ and an $N_s$  column vector $\vec h$ with entries
$h_s = H_\loc(\vec s)$.
Subtracting the sample average $O_\text{avg}$ and $h_\text{avg}$, 
\begin{equation}
\O = O - O_{\rm avg},
\qquad
\h = \vec h - h_{\rm avg},
\label{eq:Oh_avg}
\end{equation}
the QGT and the energy gradient can be approximated as
\begin{equation}
  S \approx \frac{1}{N_s}\,\O^\dagger\O,
  \qquad
  \vec g \approx \frac{1}{N_s}\,\O^\dagger\h,
  \label{eq:S_g}
\end{equation}
where the approximation reflects finite-sample statistical fluctuations.
The centering in Eq.~\eqref{eq:Oh_avg} removes the component parallel to the wavefunction, ensuring projective invariance of the quantum metric. 

The tVMC algorithm therefore consists of:
(1) sampling configurations $\vec s \sim P_{\vec\theta}$;
(2) constructing $S$ and $\vec g$;
(3) solving Eq.~\eqref{eq:sr} for $\dot{\vec\theta}$; and
(4) integrating the parameter flow on the variational manifold via a Runge-Kutta (RK) integrator~\cite{RK}.

A major challenge in tVMC is that the QGT $S$ is often severely ill-conditioned, making the linear problem in Eq.~\eqref{eq:sr} difficult to solve stably.
For ground-state searches, where the precise time trajectory is irrelevant, one may add Tikhonov regularization
$S \to S + \epsilon I$ with typically $\epsilon \sim 10^{-3}$.
However, in real-time dynamics a large $\epsilon$ introduces an unphysical bias, whereas too small a value leaves the system unstable.
In NQS-tVMC, several sophisticated regularization schemes have been proposed~\cite{nqstvmc_3, nqstvmc_7}, but none fully resolve the problem~\cite{Lange_2024}.

In PEPS, the main source of instability can be identified by the gauge redundancy, which introduces many null vectors into $O$ and hence into $S$.
The crucial feature, however, is that these null directions are \emph{analytically identifiable}.
As we will show in Sec.~\ref{sec:sr_peps}, they can be removed \emph{before} solving the TDVP equation, providing an exact preconditioning.
After this gauge-removal step, the TDVP equation becomes remarkably well conditioned, and in all examples of local quench dynamics studied we find it can be solved with exceptional numerical accuracy.
This stability is a key factor in the success of PEPS--tVMC.

\section{Stochastic reconfiguration with PEPS}
\label{sec:sr_peps}
In this section, we derive the TDVP equations for PEPS, analyze the structure of
their null space, and describe how gauge redundancy can be eliminated
efficiently in closed form, in both the TDVP and its reduced variant based on the push-through identity  (referred to as minSR) \cite{minsr}.

\begin{figure}[t]
  \centering
  \includegraphics[width=0.8\columnwidth]{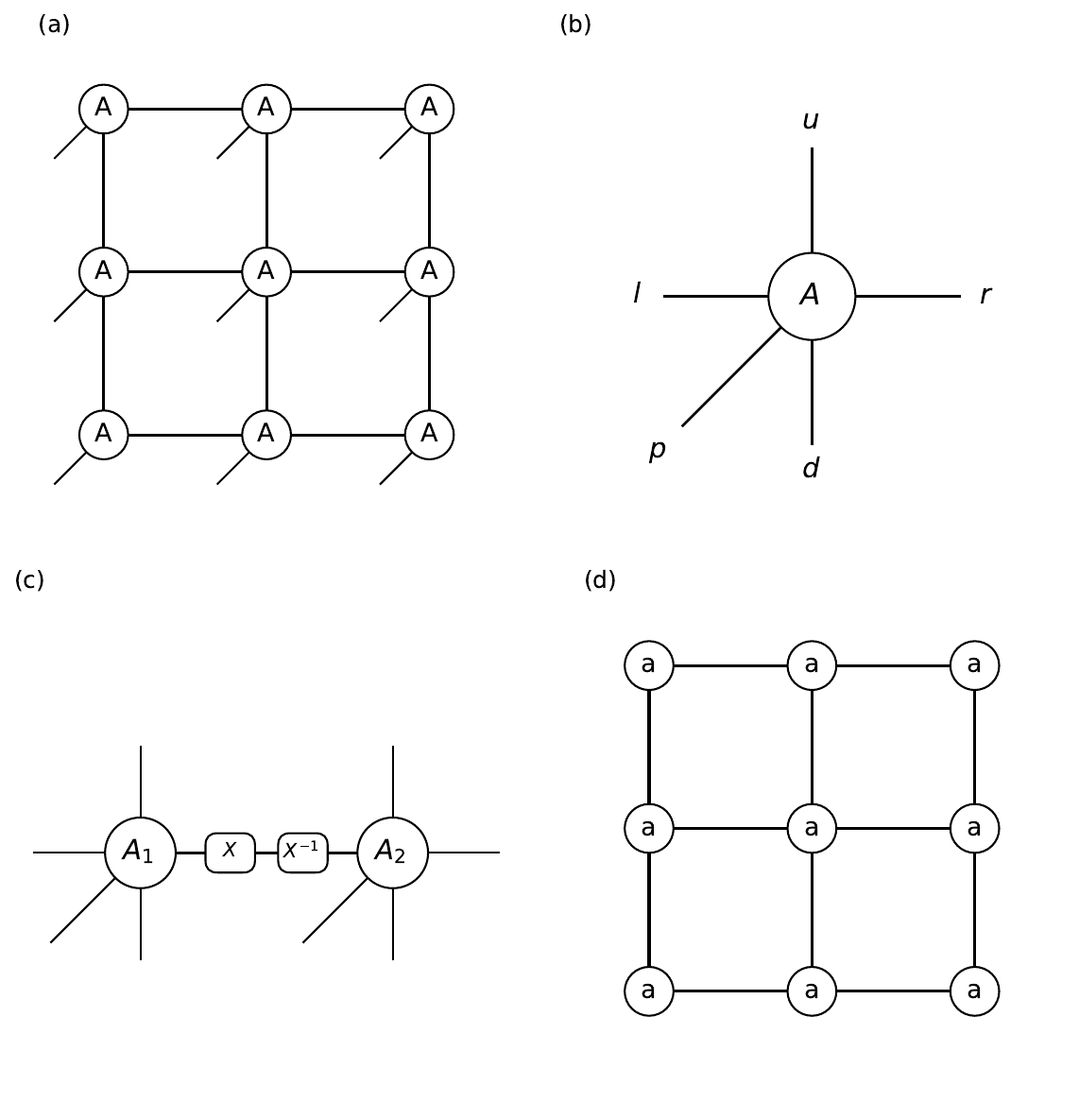}
  \caption{
    PEPS structure and gauge freedom.
    (a) A $3\times3$ PEPS network with local tensors $A$.
    (b) A single tensor $A$ with physical index $p$ and virtual indices $(l,r,d,u)$.
    (c) Local gauge transformation on a virtual bond: inserting an invertible matrix $X$ and its inverse $X^{-1}$ between two adjacent tensors leaves the global PEPS wavefunction invariant.
    (d) Wavefunction amplitude $\braket{\vec s|\Psi}$, computed via the contraction of the single layer network of sliced PEPS tensors: $a[\vec x]_{lrdu} = A[\vec x]^{p=\vec s[\vec x]}_{lrdu}$.
  }
  \label{fig:peps}
\end{figure}

A PEPS defines a variational wavefunction $\Psi_{\vec\theta}(\vec s)$ parametrized by the tensor elements
$A[\vec x]^p_{lrdu}$ at each lattice site $\vec x$, as illustrated in Fig.~\ref{fig:peps}(a,b).
Each index tuple $(\vec x,p,l,r,d,u)$ corresponds to one variational parameter, replacing the abstract index $\alpha$ introduced earlier.
The physical index $p$ labels the local Hilbert-space configuration (e.g., $p=0,1$ for vacuum or occupation), while the virtual indices $(l,r,d,u)$ take values from $1$ to $D$ (except on boundaries).
The PEPS in this paper all have open boundary conditions.
\paragraph{Gauge redundancy.}
The PEPS representation contains a large local gauge freedom associated with internal virtual bonds.
As shown in Fig.~\ref{fig:peps}(c), the transformation
\begin{align}
  {A_1}^p_{lrdu} &\rightarrow \sum_{r'=1}^D {A_1}^p_{lr'du} X_{r'r},\\
  {A_2}^p_{lrdu} &\rightarrow \sum_{l'=1}^D (X^{-1})_{ll'} {A_2}^p_{l'rdu},
\end{align}
leaves the global wavefunction invariant for any invertible $D\times D$ matrix $X$.
The infinitesimal generators of such transformations define \emph{gauge vectors}.
Each gauge vector $\vec v$ is a null vector of $O$:
\[
   \sum_\alpha O_{s\alpha}\, v_\alpha = 0 \qquad \forall\, s.
\]
Hence every gauge vector also lies in the null space of 
$S = \O^\dagger \O$, and must be removed before solving the TDVP equation.

For an $L\times L$ PEPS there are $2L(L-1)$ internal bonds, yielding $2L(L-1) D^2$ gauge vectors.
Let $E^{ij}$ be the standard basis of $\mathbb{C}^{D\times D}$ with a single nonzero element at $(i,j)$.
For the bond in Fig.~\ref{fig:peps}(c), the corresponding gauge vector $\vec v$ has components
\begin{align}
  \vec v[\vec x_1]^p_{lrdu} &= \sum_{r'} {A_1}^p_{lr'du}\,E^{ij}_{r'r},
  \label{eq:v1}\\
  \vec v[\vec x_2]^p_{lrdu} &= -\sum_{l'} E^{ij}_{ll'}\,{A_2}^p_{l'rdu},
  \label{eq:v2}
\end{align}
and vanishes for all other parameters.

\paragraph{Linear dependence of gauge vectors.}
The $2L(L-1) D^2$ gauge vectors are not linearly independent.
Consider a closed loop of virtual bonds and replace $E^{ij}$ with the identity matrix on each bond.
The sum of the corresponding gauge vectors vanishes, indicating linear dependence. 
These relations are generated by the $(L-1)^2$ elementary plaquettes of the square lattice.
We denote the associated constraint vectors by $\vec w$, which are known exactly in the basis of the gauge vectors and independent of tensor content.
\paragraph{Additional null vectors.}
In addition to gauge redundancy, two further null transformations of $\O$ exist:
(i) a global rescaling $A \to aA$, reflecting the projective nature of quantum mechanics; and
(ii) for fixed particle number $M$, a relative rescaling
$A^{p=0} \to aA^{p=0}$ and $A^{p=1} \to bA^{p=1}$ with $a^{L^2-M}b^M = 1$.
We denote their generators as $\vec u_1$ and $\vec u_2$.
Note that $\vec u_1$ is not a null direction of $O$, but only of $\O$.

\subsection{Gauge-fixing within TDVP}
We now outline the procedure for eliminating null vectors within the TDVP formulation.
First, we construct a projection matrix onto the subspace orthogonal to the constraint vectors $\vec w$, thereby reducing the number of independent gauge vectors to
\[
N_{\rm gv} = 2L(L-1)D^2 - (L-1)^2.
\]
We then form an $N_p \times (N_{\rm gv}+2)$ matrix $T$ whose columns consist of these independent gauge vectors together with $\vec u_1$ and $\vec u_2$.
Performing a \textit{complete} \texttt{QR} decomposition $T = \widetilde Q R$, we define $Q$ as the last $N_p - N_{\rm gv} - 2$ columns of $\widetilde Q$.
Since $T$ has full column rank, the remaining columns of $\widetilde{Q}$ span the orthogonal complement to the column space of $T$.
By construction, $Q$ projects onto the subspace orthogonal to all null directions.

The TDVP equation is then solved for $y$ in the reduced space:
\begin{equation}
  Q^\dag \O^\dag \O\,Q\, y = -i\,Q^\dag \O^\dag \h,
\end{equation}
and the parameter update is obtained as $\dot{\vec\theta} = Qy$.
Only a single \texttt{QR} decomposition is required, which is numerically stable and GPU efficient.

\subsection{Gauge-fixing within minSR}
When the number of samples is less than the column rank of $\O$, i.e. $N_s < N_p - N_{\rm gv} - 2$, the TDVP equation can be equivalently solved in the dual minSR form obtained using the push-through identity~\cite{minsr}:
\begin{equation}
    \mathcal{O}\, \mathcal{O}^\dagger\, y = \h,
    \qquad
    \dot{\vec\theta} = \mathcal{O}^\dagger y .
\end{equation}
We point out that minSR automatically achieves gauge removal, as the number of samples now determines the rank of the QGT matrix.
That is, tVMC provides, via stochasticity, a surprising manner to eliminate gauge redundancies, which is entirely absent in the traditional approach to tensor networks.

Assuming decorrelation between samples, the matrix $\mathcal O^\dagger$ has a single null vector $[1,1,\dots,1]^T$, which we remove with the same \texttt{QR} projection described above.
More generally, the null vector of $\O^\dag$ is the weight vector $W$ used in performing the statistical average: $\braket{A} \approx \sum_{s=1}^{N_s} W_s A_\text{loc}(s)/N_s$, which in our case is simply the identity weight.
A non-trivial weight vector would arise if reweighting is used in the Monte Carlo estimate.
The removal of this null vector may be also beneficial to NQS-tVMC in the minSR form.

\subsection{minSR using $O$ instead of $\O$}
We present a new form of minSR which does not even have the zero mode $[1,1,\cdots,1]^T$. 
\lemma{
If $\vec x$ satisfies 
\begin{equation}
O \vec x = \vec h 
\label{eq:Ox=h}
\end{equation}
then it satisfies the tVMC equation $\O^\dag \O \vec x = \O^\dag \h$. 
}
\proof{
If $\sum_{\alpha} O_{s\alpha} x_\alpha = h_s$, then 
\begin{equation}
\braket{O} \vec x = \frac{1}{N_s} \sum_s \sum_{\alpha} O_{s\alpha} x_\alpha = \frac{1}{N_s}\sum_s h_s = \braket{h} 
\end{equation}
Thus, $\O \vec x = (O - \braket{O}) \vec x = \vec h-\braket{h} = \h$.
Applying $\O^\dag$ to both sides gives the tVMC equation.\qed
}

In the TDVP regime, $N_s > N_p$, a solution $\vec x$ to Eq. \ref{eq:Ox=h} generally does not exist, so one cannot solve the TDVP equation via Eq. \ref{eq:Ox=h}. 
In the minSR regime, $N_s < N_p$ and Eq. \ref{eq:Ox=h} is under-determined, and one generally can find many solutions to it.  
Via the fundamental theorem of linear algebra: $\ker(O)^\perp = \text{im}(O^\dag)$, we pick the solution orthogonal to the kernel of $O$: $x = O^\dag y$ and solve for $y$: 
\begin{equation}
  OO^\dag y = h, \hspace{5mm} x = O^\dag y
  \label{eq:minsr_new}
\end{equation}
In contrast to the conventional minSR matrix $\O\O^\dag$, $OO^\dag$ generally does not have any zero modes at all. 
We use Eq. \ref{eq:minsr_new} to perform minSR in the examples. 

Eq. \ref{eq:Ox=h} also has a nice interpretation. 
It can be ``derived'' directly from the Schrödinger equation. 
If quantum states are parametrized with variational parameters $\theta_\alpha$ and satisfies the Schrödinger equation, then for \textit{any} configuration $\vec s$,
\begin{equation}
 \sum_{\alpha} \braket{\vec s|\partial_\alpha\Psi} \delta \theta_\alpha =  -i\delta t \braket{\vec s|H|\Psi}
  \label{eq:Schrödinger}
\end{equation}
Dividing by $\braket{\vec s|\Psi}$ gives one row of Eq. \ref{eq:Ox=h} where $\vec x = i\vec{\dot{\theta}}$, for sample $\vec s$.
Thus, the Schrödinger equation requires Eq. \ref{eq:Schrödinger} for every configuration $\vec s$, while Eq. \ref{eq:Ox=h} requires it for a stochastically selected set of configurations.

\subsection{Small-$o$ trick in minSR}
In the minSR case, a small-$o$ trick can be specifically applied to PEPS to alleviate the memory cost in storing the full matrix $O$.
The key observation is that the \emph{locality} of PEPS forces most elements $O_{s\alpha}$ to vanish.
For a given sample configuration $\vec s$,
\begin{equation}
    O_{s\alpha}
    =
    \frac{\partial\log\Psi(\vec s)}
         {\partial A[\vec x]^p_{lrdu}}
    = 0
    \quad \text{if } p \neq s(\vec x).
    \label{eq:smallo_0}
\end{equation}
Thus each site contributes only the physical slice compatible with the sample.
We therefore define the ``small-$o$'' object
\begin{equation}
    o[\vec x]_{lrdu}(\vec s)
    \equiv
    O[\vec x]^{\,p = s(\vec x)}_{lrdu}(\vec s),
\end{equation}
from which the full minSR matrix $OO^\dagger$ can be reconstructed with the memory cost of $o$ instead of $O$ (see the Supplementary Material (SM) \cite{SM} for the explicit GPU-efficient construction).

In tVMC, the memory bottleneck is storing $O$ or $S$, which can be a serious issue on GPUs.
For a PEPS with local Hilbert space dimension $d$, this ``small-$o$'' trick reduces memory cost by a factor of $d$.
For the $\mathbb{Z}_N$ gauge-invariant PEPS used in Sec.~\ref{subsec:lgt}, the reduction is very substantial—a factor of $dN^3$.
In practice, this saving often determines whether a large-scale GPU simulation is feasible at all.

\subsection{TDVP versus minSR}
We comment on the choice between TDVP and minSR form.
As the system size increases, $N_p$ eventually overtakes $N_s$ and one should use minSR for efficiency.
However, it is always advised to study a problem first at small size and bond dimension and ideally benchmark against exact methods, in which case typically $N_s > (N_p - \text{gauge redundancy})$ and it is only correct to use standard TDVP.
In fact, the examples we will present are all benchmarked first against exact diagonalization at small system sizes. 
Thus, the TDVP form and its explicit gauge removal is very important in the workflow. 
Even though the results presented may be of large system size simulated with minSR, many calculations were done in the TDVP form before that.

\subsection{Comparison with double-layer methods}
\label{subsec:comparison}
We briefly discuss the difficulties of double-layer methods mentioned in the introduction: computational cost, gauge fixing, and positivity of the quantum metric.
Although manifested differently, they reflect the same underlying difficulty.
\paragraph{Computational cost.}
Consider a PEPS with bond dimension $D$.
Let $D'$ denote the bond dimension of the boundary MPS (bMPS) used to contract a single-layer PEPS in VMC, cf. Fig. \ref{fig:peps} (d), and likewise the boundary matrix product operator (bMPO) used in full-update double-layer contraction.
The dominant computational costs scale as $D^4 D'^2$ for VMC~\cite{liu2021} and $D^4 D'^3$ for full update (FU)~\cite{algorithm_finite_peps}.
Typical choices are $D' \sim D$ to $D^2$ for VMC, and $D' \sim D^2$ to $D^3$ for full update, leading to effective complexities of $D^6$–$D^8$ and $D^{10}$–$D^{13}$, respectively.
This gap is substantial.
VMC additionally incurs sampling overhead, but even choosing $D'$ larger than $D^3$ does not reliably stabilize the full-update procedure due to the next two issues.
\paragraph{Gauge fixing.}
  Unlike MPS in 1D~\cite{cirac2021matrix}, PEPS does not possess a canonical form that fixes the gauge freedoms.
  Ref.~\cite{algorithm_finite_peps} introduces local gauge fixing for full-update gate applications,
but no globally consistent gauge-fixing procedure is known.
  In fact, as an alternative, a variational subfamily --- isometric PEPS (isoPEPS) \cite{zaletel2020isometric,lin2021efficient,dai2025fermionic} --- is introduced to explicitly preserve the isometric gauge at the expense of expressiveness.  
 However, the necessary shifting of isometric centers introduces significant errors and leads to suboptimal results. 
 In contrast, PEPS–tVMC achieves global gauge fixing efficiently through a single \texttt{QR} projection or simply using the minSR formulation, without any loss of representational power.
 \paragraph{Positivity.}
 Because PEPS are not in an isometric form, to perform a variational update, one needs to also form the non-trivial quantum metric of the parameters, induced from the inner product of the Hilbert space.  
This quantum metric is also known as the norm matrix $N$ in Ref. \cite{algorithm_finite_peps}. 
In practice, approximate contraction leads to violations of positivity of $N$, which destabilizes the generalized eigenvalue problems used in variational updates, especially in time evolution.
 Tiny errors in the boundary contraction can lead to large deviations in the time trajectory. 
 tVMC avoids this issue entirely: normalization is enforced simply by subtracting $O_{\rm avg}$ and $h_{\rm avg}$ in Eq.~\ref{eq:Oh_avg}, and the $S$ matrix in Eq. \ref{eq:S}, serving as the quantum metric, is manifestly positive semi-definite.
\paragraph{Abelian symmetry.}
In VMC, any conservation law that is diagonal in the computational basis can be directly implemented by sampling configurations in the target charge sector.
This is equivalent to acting on the state with a global projector $P_M$ onto the space with desired total charge $M$: 
\begin{equation}
  \ket{\Psi} = P_M \ket{\text{PEPS}_{\vec \theta}}
\end{equation}
where only the underlying PEPS is parametrized.
If written out explicitly, the global projector $P_M$ is very expensive: it is a snake MPO with a bond dimension linear to system size. 
The VMC framework, however, allows such a projector not to incur any computational or memory cost, significantly increasing the representability of the underlying bare PEPS.

\section{Results}
We now apply PEPS–tVMC to study the real-time dynamics of six distinct two-dimensional models—bosonic and fermionic, free and interacting, and including systems with gauge symmetry or nontrivial topology. 
PEPS configurations are sampled using the sequential sampling algorithm, which was first introduced in Ref. \cite{liu2021}, and later proved to satisfy detailed balance in Ref. \cite{wu2025algorithm}. 
The local energies and log-derivatives are computed via the contraction of the single-layer network shown in Fig. \ref{fig:peps}(d) (see Refs. \cite{liu2021,wu2025algorithm,wu2025accurate} for methodological details). 
For Hamiltonians with no explicit time-dependence, the energies are observed to be well-conserved, with relative drift of $2\times10^{-4}$ to $2\times10^{-3}$ for the entire trajectory.
We show the energy evolution in the Supplementary Material \cite{SM}.
The examples are all on squares with open boundary conditions.
In examples \ref{subsec:chern}-\ref{subsec:sf}, the initial states of the dynamics are all obtained with ground state PEPS VMC with the same $D$ and $D'=3D$ for the Chern insulator and the superfluid, $4D$ for the Hofstadter model, and $3\frac{D}{2}$ for the $\Z_2$ lattice gauge theories. 
\label{sec:results}
\subsection{Free Fermion Chern insulator}
\label{subsec:chern}
As a first test, to obtain nontrivial exact benchmark, we consider a $12\times12$ free-fermion Hofstadter model with flux 
$\phi=2\pi/3$,
\begin{equation*}
  H_\text{chern} = -\sum_{x, y} c_{x,y}^\dag c_{x+1,y} + e^{-i\frac{2\pi x}{3}} c_{x,y}^\dag c_{x,y+1} + \text{h.c.}
\end{equation*}
We fix the particle number to $2/3$ filling, populating two of the three magnetic subbands.
With open boundaries, the system hosts gapless chiral edge modes.
Using fermion PEPS \cite{fPEPS_VF, fPEPS_SG, wu2025algorithm}, we prepare the ground state in a weak attractive potential (strength $-1$) placed at the upper-left corner, which creates a localized edge-density bump. 
The system is then evolved in real time under $H_{\rm chern}$ with the potential removed.
\begin{figure}[hbt]
    \centering
    \includegraphics[width=\linewidth]{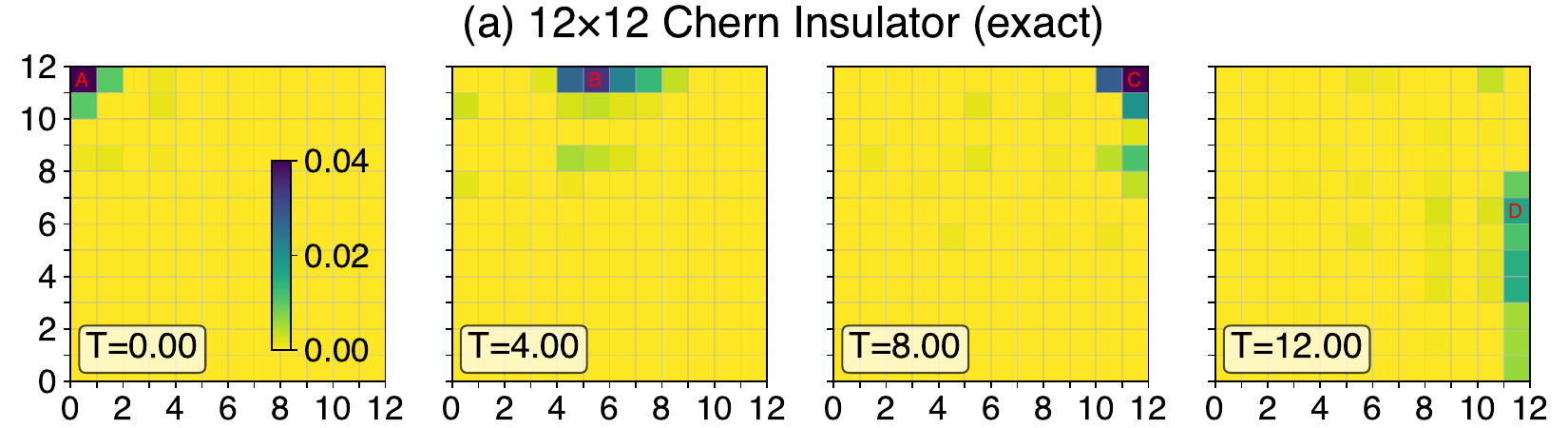}
    \includegraphics[width=\linewidth]{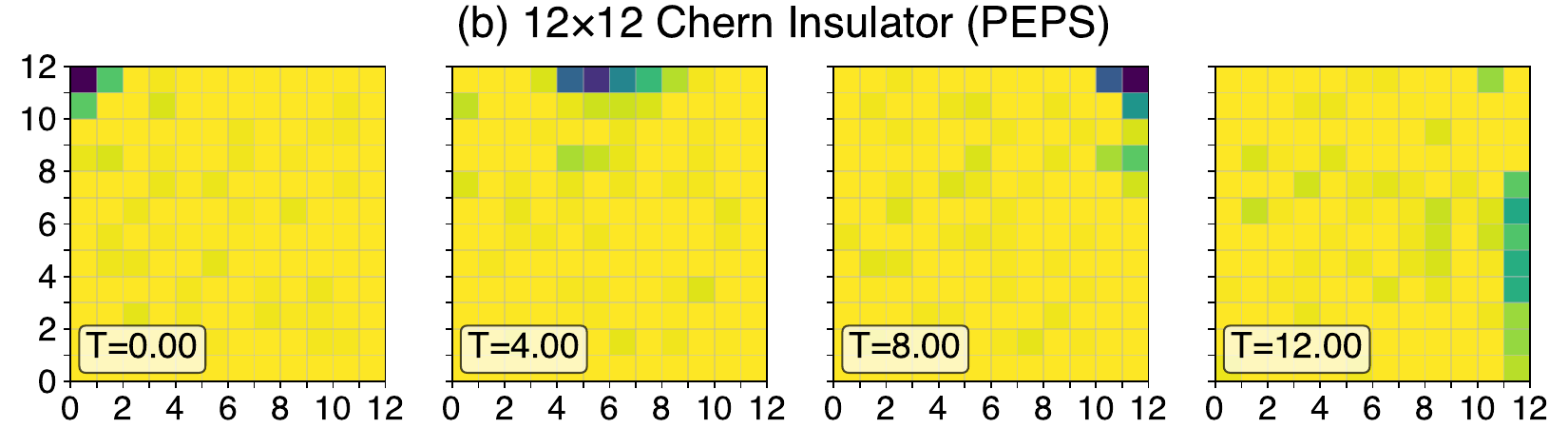}
    \includegraphics[width=\linewidth]{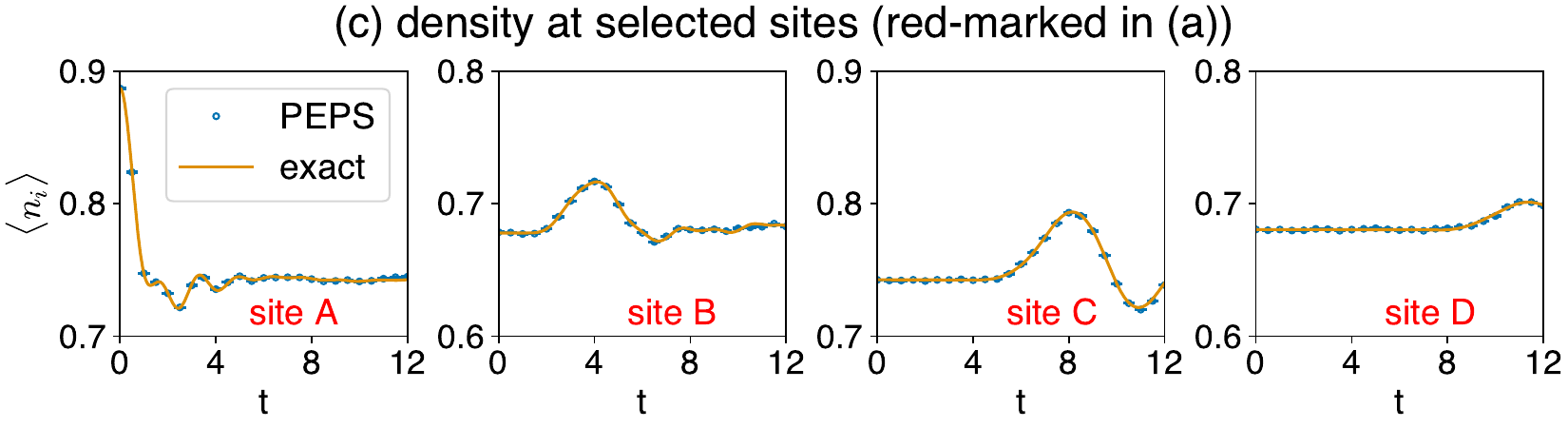}
    \includegraphics[width=\linewidth]{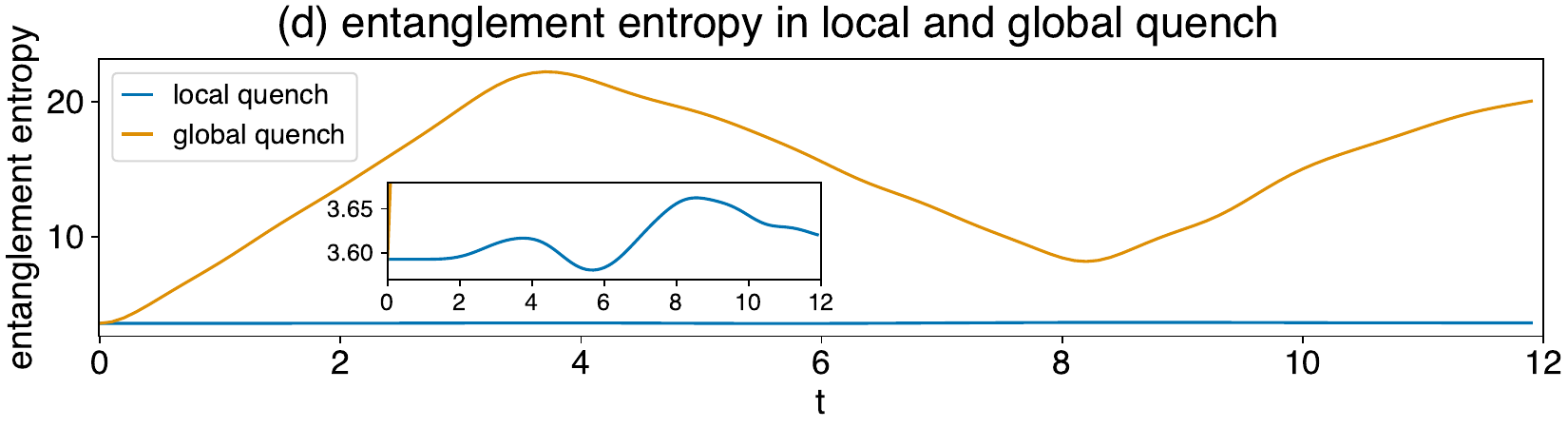}
    \caption{
    Chiral edge dynamics in a $12\times12$ Chern insulator.
    (a) Exact real-time evolution obtained from free-fermion numerics. The color scale shows the density deviation $\braket{n_i(t)} - \braket{n_i}_\text{gs}$.
    A density perturbation created near the boundary at $T=0$ propagates unidirectionally along the edge at later times $T=4, 8, 12$.
    (b) Same protocol computed using PEPS--tVMC.
    (c) 
    Time evolution of the density at four selected sites red-marked in (a), chosen so that the chiral pulse passes them at $T=0, 4, 8, 12$. 
    The error bars are of order $7\times10^{-4}$. 
    All panels share the same vertical scale of 0.2. 
    (d) Comparison of the entanglement entropy of the left half of the system during local and global quench dynamics, computed with free fermion numerics.
    The local quench is the same as in (a). 
    The global quench is the dynamics under $H_\text{chern}$ with magnetic field removed.
    }
    \label{fig:chern}
\end{figure}

As shown in Fig.~\ref{fig:chern}, the excitation propagates unidirectionally along the edge.
PEPS--tVMC faithfully reproduces the edge motion, its speed, and its spatial profile, in excellent agreement with the exact free-fermion numerics.
For context, isoPEPS simulations of a similar chiral edge motion (e.g., Fig.6(b) of Ref.\cite{dai2025fermionic}) already provided state-of-the-art results at the time; however, their accuracy was ultimately limited by the errors due to the shifting of isometric centers.
The present PEPS–tVMC approach achieves noticeably higher accuracy. 
We further note that the numerical difficulty of tensor network simulations is largely dictated by the smaller of $L_x$ and $L_y$. 
Thus, although the isoPEPS study used a $9\times18$ lattice, the $12\times12$ system considered here is in fact the more challenging geometry, making this benchmark particularly stringent. 

We emphasize that the PEPS used in tVMC is entirely agnostic to the Gaussianity in the free fermion state.
Thus, despite the model being noninteracting, its non-trivial dynamics makes a demanding test for PEPS, especially at this lattice size and time scale.

In the end, taking advantage of the free-fermion numerics, we compute the entanglement entropy of the left half system during the dynamics. 
We compare it to the entanglement growth in a global quench from the same initial state, but evolving under $H_\text{chern}$ with the magnetic flux removed, i.e. all hoppings being real and positive.
As seen in Fig. \ref{fig:chern} (d), the entanglement growth in the global quench is much faster, peaking at around $t \approx 3.8$ due to the entanglement quasi-particle \cite{Calabrese2005Evolution} hitting the edge of the system.
The local quench we simulate has a mild entanglement growth and is thus more suitable for simulation with PEPS, as discussed in the introduction.

\subsection{Pure $\Z_2$ lattice gauge theory}
\label{subsec:pureZ2}
Next, to highlight the improvement due to the analysis above, we revisit our exploratory calculation in Ref. \cite{wu2025accurate}, which solved the TDVP equation using conjugate-gradient without any gauge removal. 
We considered a pure $\Z_2$ gauge theory on a $10\times 10$ lattice: 
\begin{equation}
  H_{\Z_2} = - 2\sum_p B_p + 2g\sum_l (1-Z_l)
\end{equation}
subject to the Gauss-law constraint $\prod_{l\in i} Z_l = 1$.
Here $B_p=\prod_{l\in p}X_l$ measures the magnetic flux through plaquette $p$, and a violation of $B_p=1$ defines a \textit{vison} excitation \cite{vison}.
$X_l, Z_l$ are the Pauli $x, z$ matrices for the gauge field on link $l$.
We consider the deconfined phase with $g = 0.1$, where the visons are the elementary excitations of the system and are expected to propagate with a finite life-time.  

\begin{figure}[hbt]
    \centering
    \includegraphics[width=\linewidth]{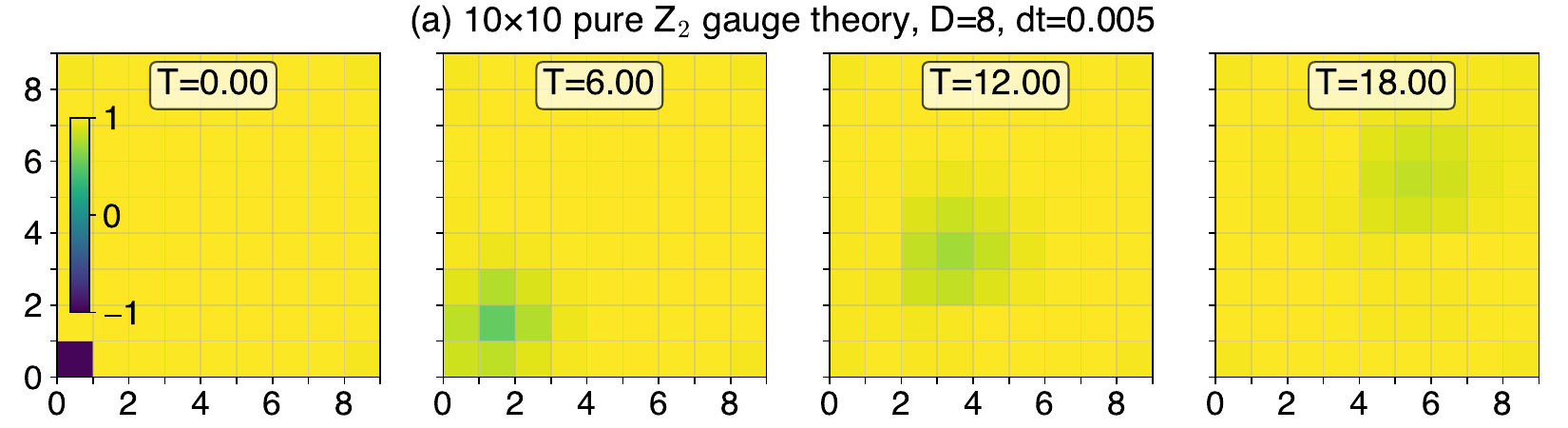}
    \includegraphics[width=\linewidth]{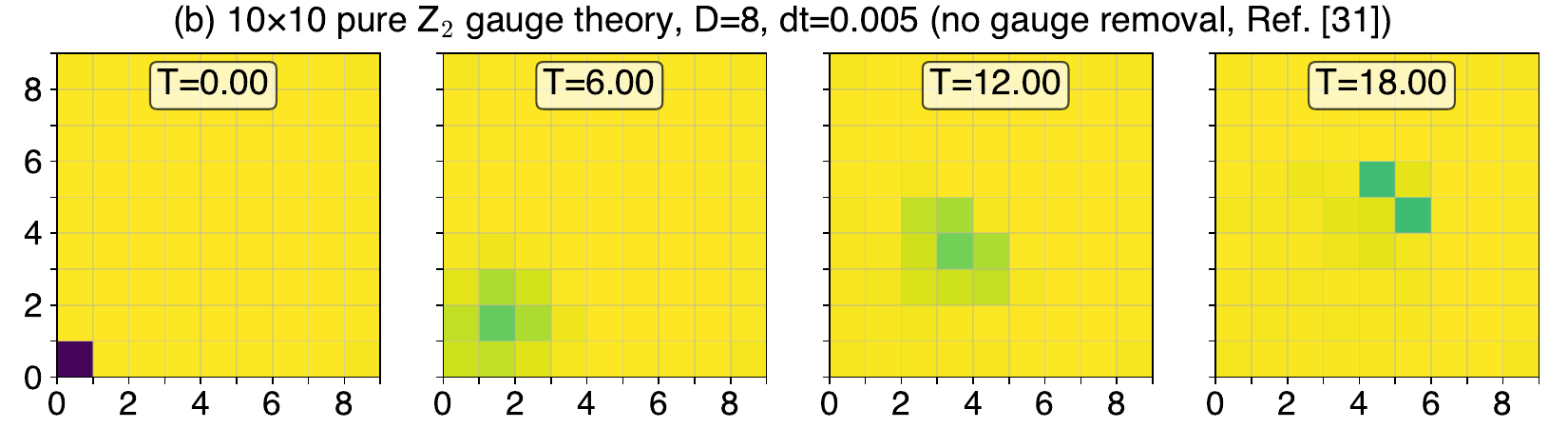}
    \includegraphics[width=\linewidth]{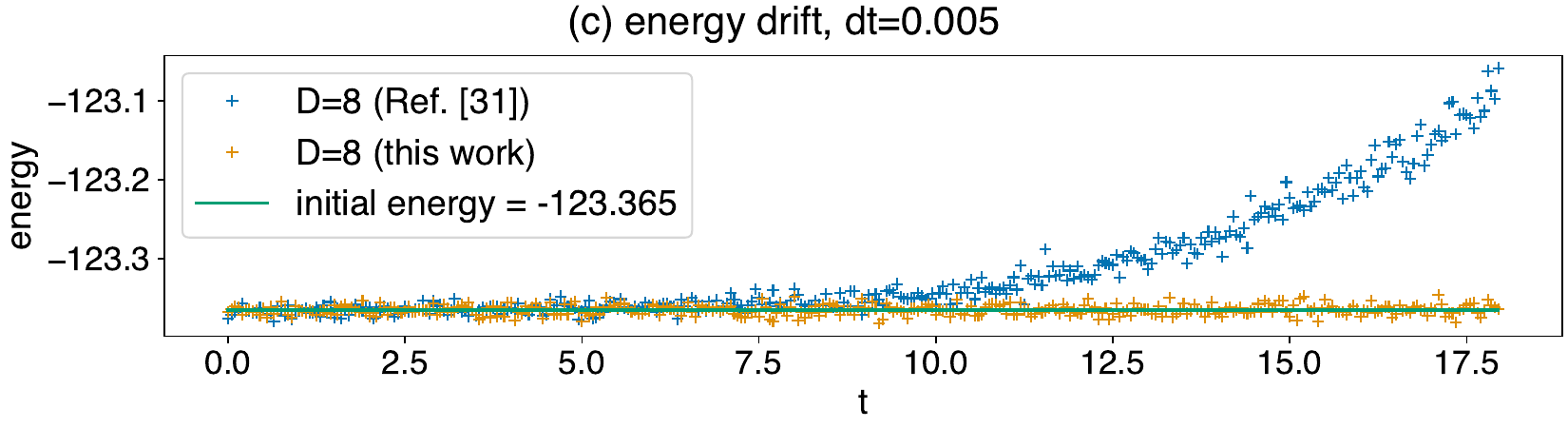}
    \includegraphics[width=\linewidth]{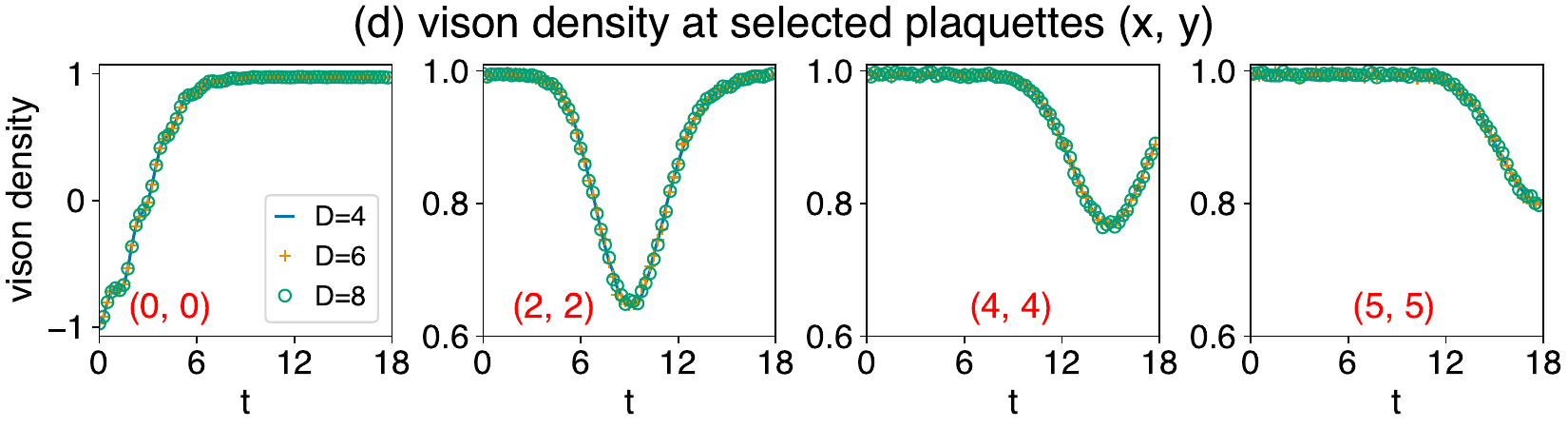}
    \caption{
Real-time vison propagation in the pure $\Z_2$ lattice gauge theory with $g=0.1$ on a $10\times10$ lattice.
(a) Snapshots of the plaquette operator $B_p(t)$, which highlights magnetic-flux (vison) excitations created at $T=0$, simulated with gauge removal. 
(b) Data reported in Ref. \cite{wu2025accurate}.
(c) Energy evolution during the simulations.
(d) Vison densities at various bond dimensions.
    }
    \label{fig:pureZ2}
\end{figure}

Following the same setting as in Ref. \cite{wu2025accurate}, we create a vison at the lower-left corner of the lattice by acting on the ground state an $Z_l$ operator on the link right to the site $(0, 0)$.   
We then simulate the subsequent tVMC dynamics under $H_{\Z_2}$ with gauge-invariant PEPS, using the same bond dimension ($D=8$), sample size ($N_b = 10240$), time step ($dt=0.005$), and integrator (2nd order Runge-Kutta) as in Ref. \cite{wu2025accurate}. 
Here, we solve the TDVP equation with Cholesky decomposition and gauge removal discussed above. 
As seen in Fig. \ref{fig:pureZ2} (c), with the current method, the energy conservation is significantly better and the vison propagation also looks more physical. 
Fig. \ref{fig:pureZ2} (a) and (b) reveal a noticeable difference in the vison densities at late time, highlighting the importance of gauge removal in PEPS-tVMC.

In fact, the entanglement in this dynamics is quite low, and the simulations are well-converged already at $D=4$.
We demonstrate the convergence in vison density with $D=4,6,$ and $8$ in Fig. \ref{fig:pureZ2} (d).
We note that the gauge-theory PEPS has a subtle structure \cite{wu2025accurate} and the computational cost for $\Z_N$ gauged PEPS with bond dimension $D$ is equivalent to the that of a normal PEPS with bond dimension $\frac{D}{N}$. 
Thus, the $D=4,6,8$ in Fig. \ref{fig:pureZ2} only incurs computational cost of $D=2,3,4$ in the normal case.

\subsection{$\Z_2$ lattice gauge theory coupled to Higgs field}
\label{subsec:lgt}
Next, for a much more challenging gauge theory dynamics, using the gauge-invariant PEPS--VMC framework introduced in Ref. \cite{wu2025accurate}, we examine the confinement dynamics in the $\Z_2$ lattice gauge theory coupled to a Higgs field,
\begin{equation*}
  H_\text{lgt} = -\sum_i \sigma_i^z - \sum_p B_p - J \sum_{l} \sigma^x_{l_-}X_l\sigma^x_{l_+} - g \sum_{l} Z_l
\end{equation*}
subject to the Gauss-law constraint $\sigma_i^z \prod_{l\in i} Z_l = 1$.
$\sigma^x_i, \sigma^z_i$ are the Pauli matrices on the site $i$, representing the Higgs field.
$l_{\pm}$ are the sites at the ends of link $l$.
The system has three ground state phases: confined (small $J$ and large $g$), Higgs (large $J$ and small $g$), and deconfined (small $J$ and small $g$) phases \cite{fradkin1979phase}.
When $g = 0$, the phase transition point between the deconfined phase and the Higgs phase is at $J_c \approx 0.328$ \cite{wu2012phase}.
In the Higgs phase, the visons are expected to have a linearly attractive interaction, i.e. vison confinement. 
In the deconfined phase, the visons are expected to propagate freely, i.e. vison deconfinement.

\begin{figure}[hbt]
    \centering
    \includegraphics[width=\linewidth]{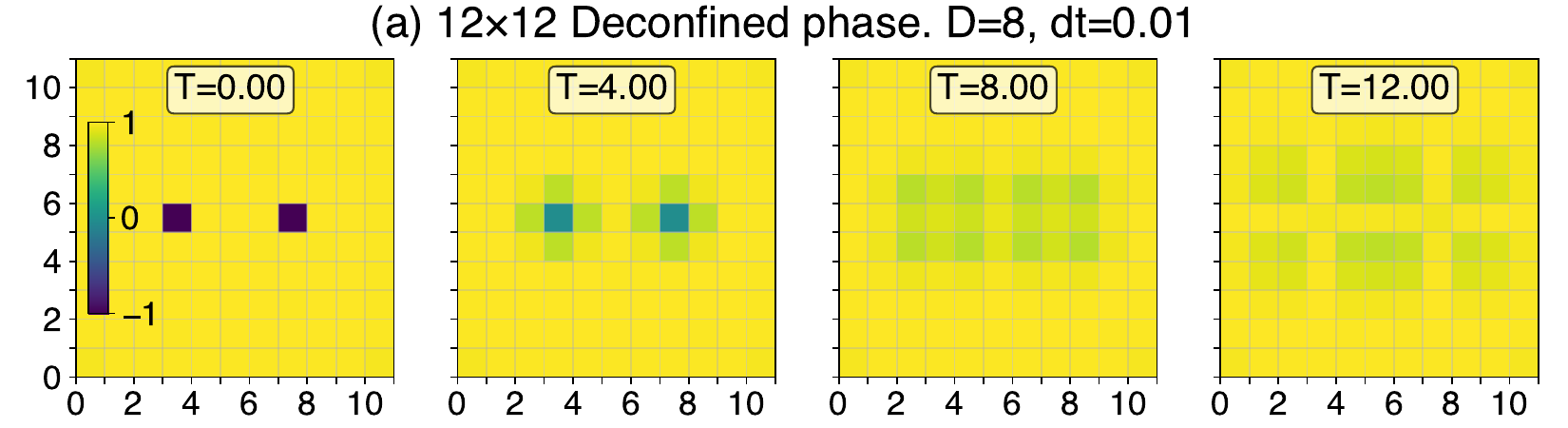}
    \includegraphics[width=\linewidth]{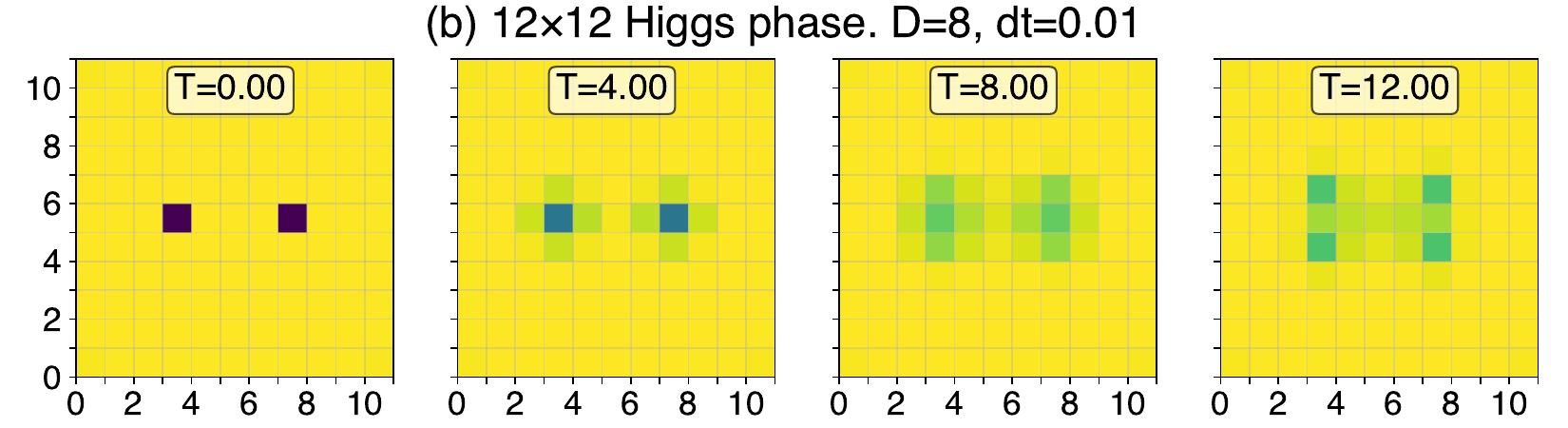}
    \includegraphics[width=\linewidth]{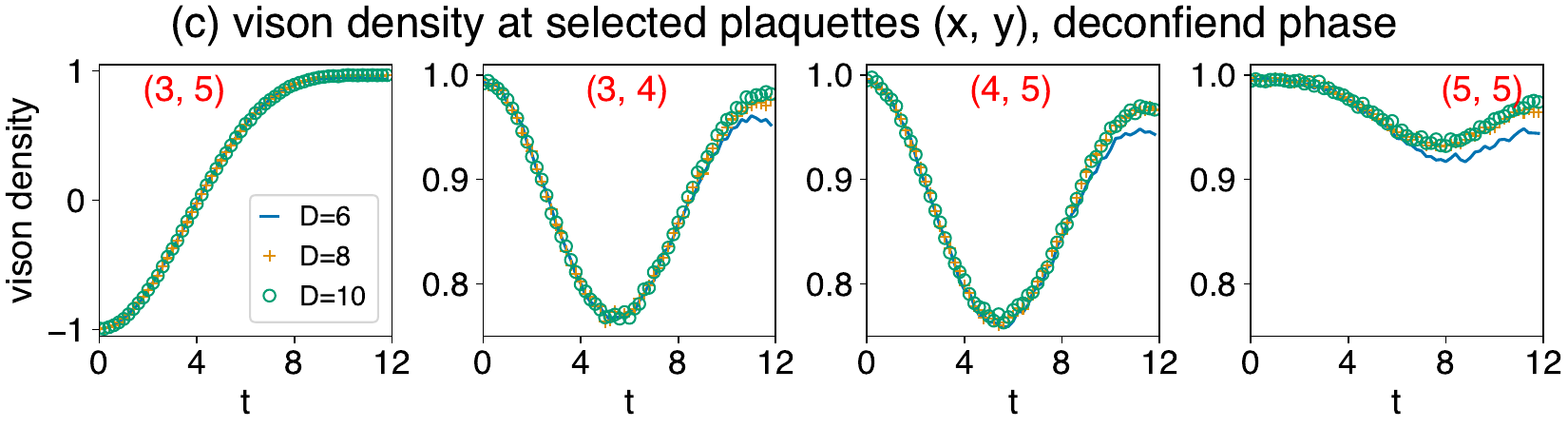}
    \includegraphics[width=\linewidth]{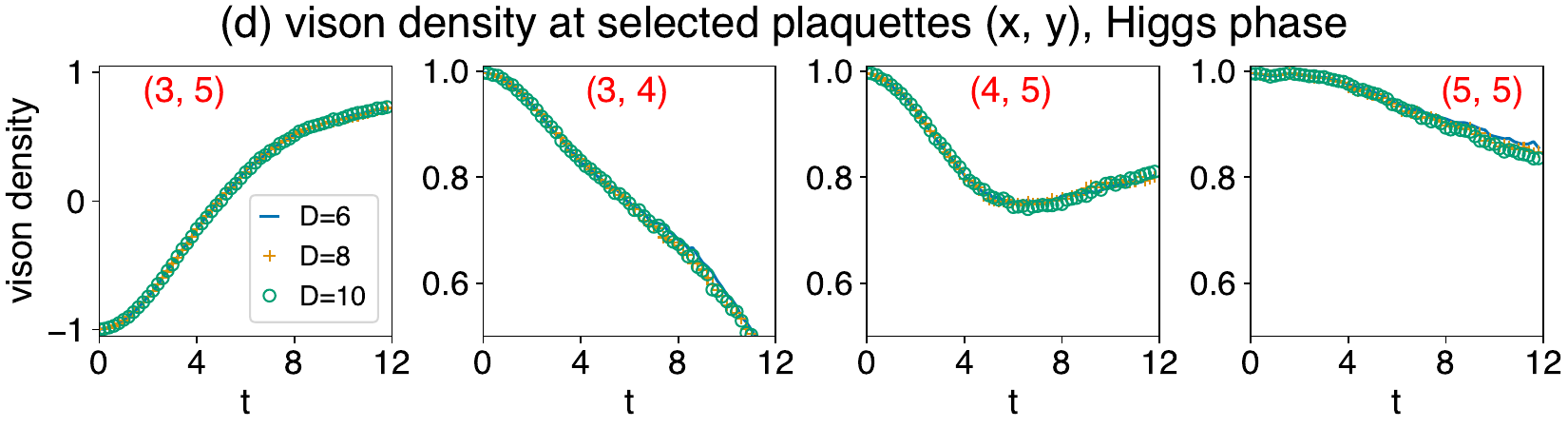}
    \caption{
Real-time vison dynamics in the $\Z_2$ lattice gauge theory on a $12\times12$ lattice.
Shown are snapshots of the plaquette operator $B_p(t)$, which highlights magnetic-flux (vison) excitations created at $T=0$.
    (a) Deconfined phase ($J=0.1, g=0.1$).
    (b) Higgs phase ($J=0.5, g=0.1$).
    (c, d) Vison densities at various bond dimensions.
    }
    \label{fig:lgt}
\end{figure}

We create a pair of visons on top of the ground state in both the Higgs and the deconfined regimes, and simulate the subsequent dynamics under $H_\text{lgt}$, shown in Fig. \ref{fig:lgt}.  
In the deconfined phase, the initially localized visons rapidly spread into extended flux clouds that delocalize across the lattice, consistent with free, deconfined excitations and the absence of a confining string.
In the Higgs phase, the visons remain bound. 
The absence of spreading reflects a finite string tension and the confining nature of the Higgs phase.
The vison densities, converged at $D=8$ and $10$, are shown in Fig. \ref{fig:lgt} (c, d) with different bond dimensions.

Recently, considerable progress has been made in implementing real-time gauge-theory dynamics on quantum devices \cite{Martinez2016,yang2020observation,Meth2025}. 
Our results highlight that classical tensor-network approaches can serve as a highly effective and complementary platform for these studies.
It would be interesting to explore whether PEPS--tVMC can probe the intricate dynamics of glueballs in the (2+1)D Yang-Mills theory, after the gauge-invariant PEPS--VMC \cite{wu2025accurate} is extended to the non-Abelian case.

\subsection{Bosonic fractional quantum Hall effect}
\label{subsec:fqhe}
Next we study the remarkable phenomena of fractional quantum Hall effect \cite{fqhe,laughlin}.
We consider a bosonic Hofstadter model with flux $\phi=\pi/2$ and hardcore constraint, on a $12\times 12$ lattice
\begin{equation*}
  H_\text{fqhe} = -\sum_{x, y} b_{x,y}^\dag b_{x+1,y} + e^{i \frac{\pi x}{2}} b_{x,y}^\dag b_{x,y+1} + \text{h.c.}
\end{equation*}
We consider the $1/8$ density, where the lowest magnetic band is half filled, corresponding to filling $\nu=1/2$. 
With the filling belonging to the Jain sequence \cite{jain1989composite}, composite fermion theory \cite{sorensen2005fractional,moller2009composite} predicts that the ground state is a bosonic analogue of the Laughlin state \cite{laughlin} with quasi-particle charge $1/2$.
The lattice structure \cite{2011fci} also complicates the theory which was initially developed for the continuum. 
We now explicitly check the fractional transport in real-time. 

\begin{figure}[hbt]
    \centering
    \includegraphics[width=\linewidth]{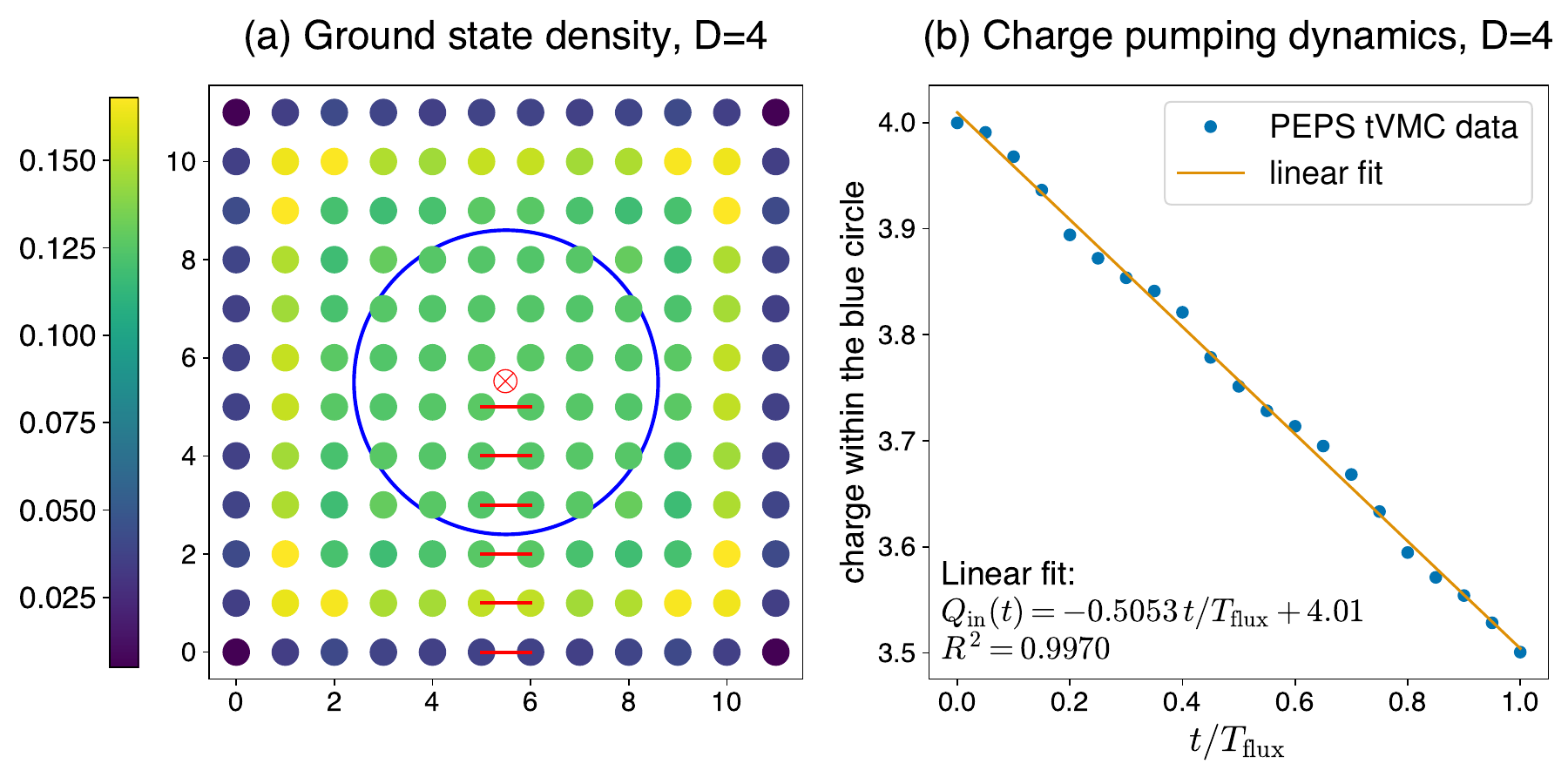}
    \includegraphics[width=\linewidth]{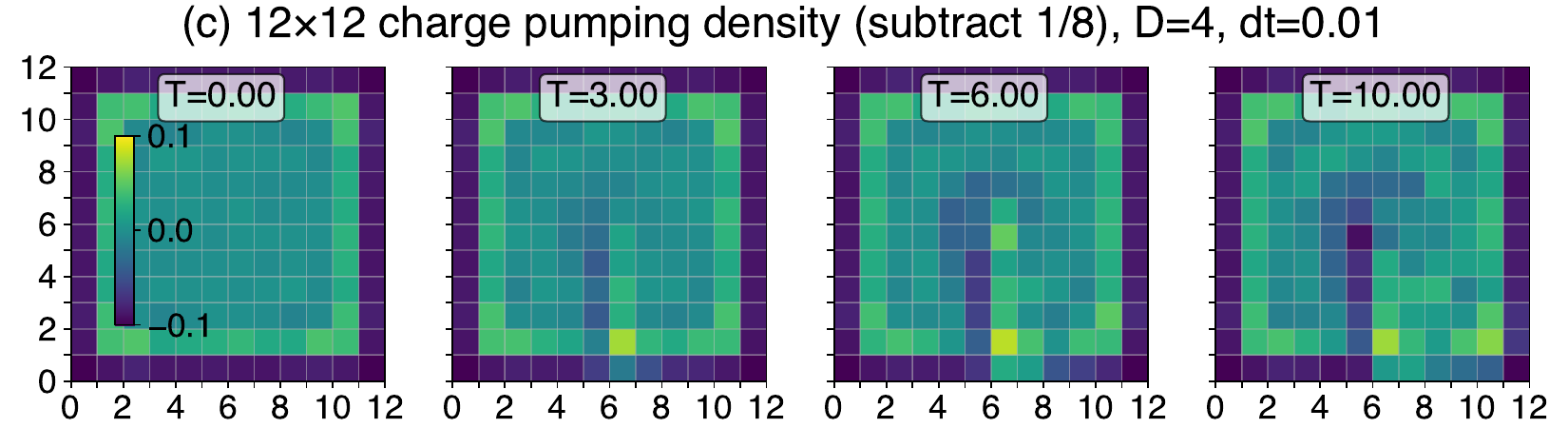}
    \includegraphics[width=\linewidth]{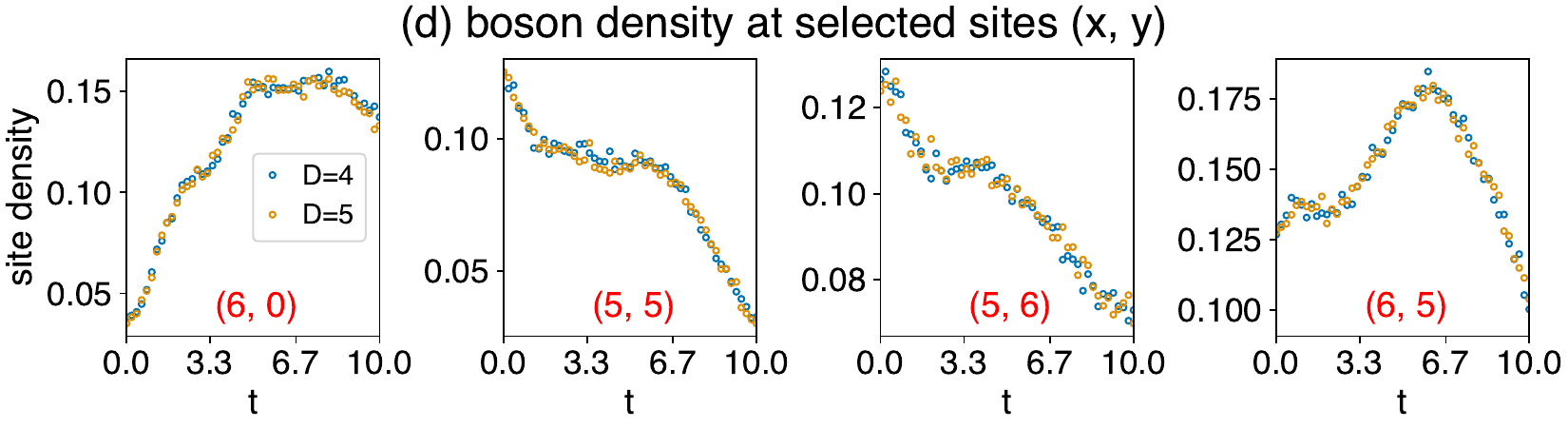}
    \caption{
Charge pumping in the bosonic Hofstadter model at flux $\phi=\pi/2$. 
(a) Ground-state charge distribution on a $12\times12$ lattice with total particle number 15.
A localized external flux is threaded through the central plaquette (marked by $\otimes$).
The red horizontal bonds indicate the Peierls-phase path along which the time-dependent vector potential $A_{ij}(t)$ is ramped.
The blue circle denotes the region over which the enclosed charge $Q_\text{in}$ is monitored.
(b) Total charge within the blue circle during one flux-insertion cycle of duration $T_\text{flux} = 10$. 
(c) Evolution of real-space density, subtracted by $1/8$, during the pump.
(d) Boson densities at various bond dimensions.
    }
    \label{fig:fqhe}
\end{figure}

To account for the finite boundary, in practice, we use particle number $(L_x-1)(L_y-1)/8 \approx 15$ \footnote{We thank Bartholomew Andrews for bringing this practice to our attention.}. 
As shown in Fig. \ref{fig:fqhe} (a), the ground state VMC gives a charge distribution that is uniformly $1/8$ in the bulk.

To probe the fractional charge transport, we thread an external time-dependent flux $\Phi(t)$ through the central plaquette, cf. the red $\otimes$ in Fig. \ref{fig:fqhe} (a), increasing linearly from 0 to $2\pi$ over a period $T_{\rm flux} = 10$.
The external flux is implemented via a Peierls phase on the red horizontal bonds shown in Fig. \ref{fig:fqhe} (a).
This generates a steady electromotive force around any closed loop enclosing the plaquette and induces a Hall transport from the middle of the system to the edge.
If the system is indeed a fractional quantum Hall state, a quantized charge of $\Delta Q = \nu = 1/2$ is expected to be transported per pumping cycle \cite{laughlin1981quantized,thouless1983pump}.
Indeed, Fig. \ref{fig:fqhe} (b) shows that PEPS--tVMC confirms this quantized transfer of charge. 
A linear fit of the charges within a central region, cf. the blue circle in Fig. \ref{fig:fqhe} (a), versus $t/T_\text{flux}$ gives a slope of $0.5053$, close to the expected Hall response of $1/2$.  

The density profile during the pumping is shown in Fig. \ref{fig:fqhe} (c). 
Because the flux is time-dependent, different Peierls paths are not gauge-equivalent at the level of time evolution, and our choice breaks the $C_4$ symmetry of the lattice.
Although the instantaneous density flow seems structureless, the integrated pumped charge remains amazingly quantized, attesting to the robustness of the Hall response to microscopic details of the system.
The boson densities are converged with bond dimension $D=4$ and $5$, as shown in Fig. \ref{fig:fqhe} (d).

This result provides a direct real-time confirmation of the composite-fermion prediction of fractional quasiparticle charge. 
While previous numerical studies of fractional Chern insulators relied primarily on static diagnostics, such as entanglement spectra or density profiles of pinned quasiholes, PEPS--tVMC enables us to directly visualize the transport of fractional charge in two dimensions.
This establishes a dynamical counterpart to the usual static
signatures of topological order.
It would be exciting to study anyonic dynamics in other settings with PEPS--tVMC in the future. 
\subsection{Superfluid bosons}
\label{subsec:sf}
Next we study real-time flow in a superfluid of hardcore bosons on a $13\times13$ lattice,
\begin{equation}
  H_\text{sf} = - \sum_{\braket{ij}} b_i^\dag b_j
\end{equation}
at half-filling.
This system is known to be in the superfluid phase due to Bose-Einstein condensation.
Although its static properties are well understood from quantum Monte Carlo \cite{Sandvik1999boson,Heinrichs1998}, its real-time dynamics—including the onset of dissipation and the critical velocity—remain much less explored, especially in two dimensions.

\begin{figure}[hbt]
    \centering
    \includegraphics[width=\linewidth]{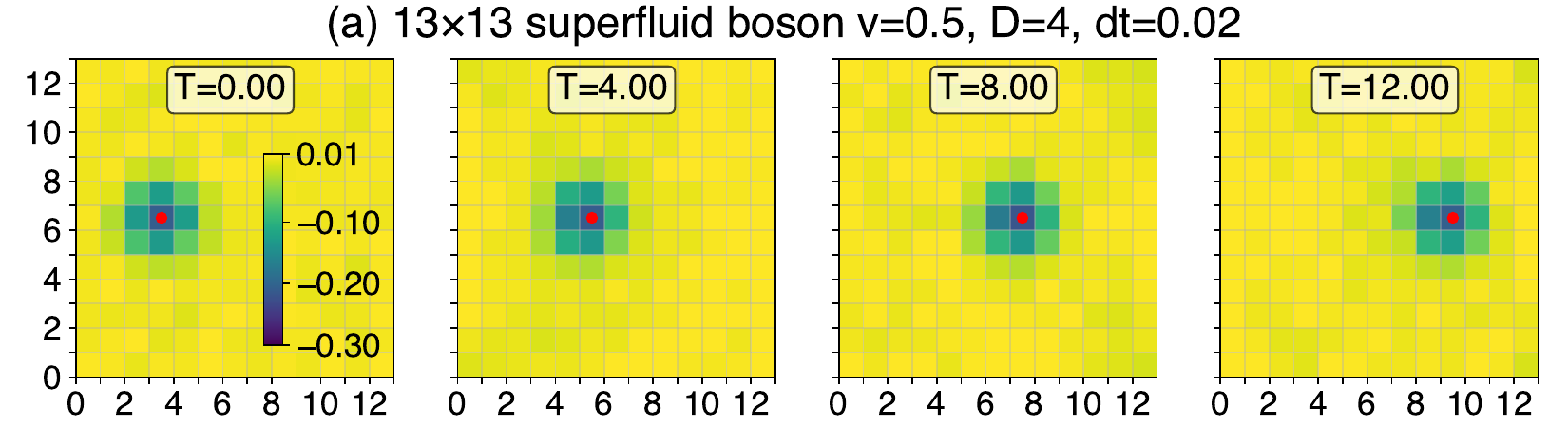}
    \includegraphics[width=\linewidth]{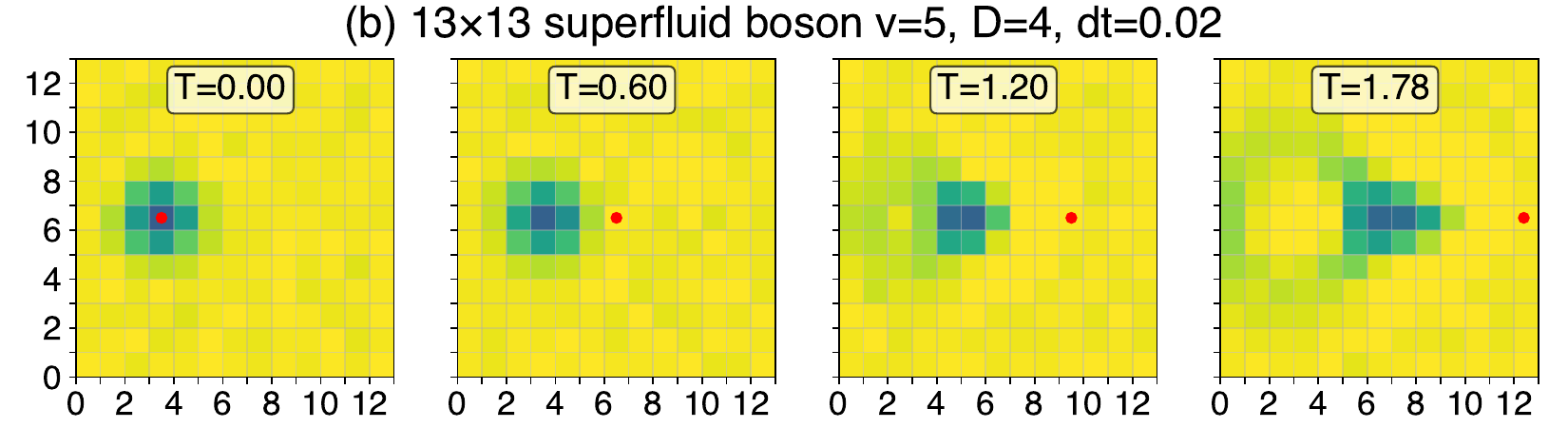}
    \includegraphics[width=\linewidth]{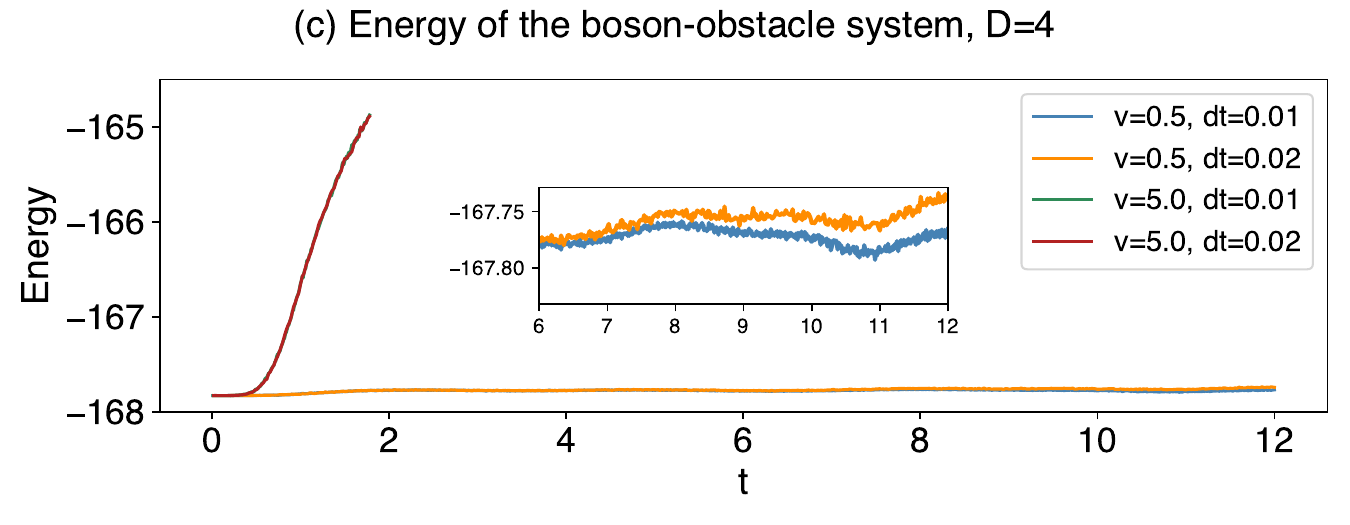}
    \includegraphics[width=\linewidth]{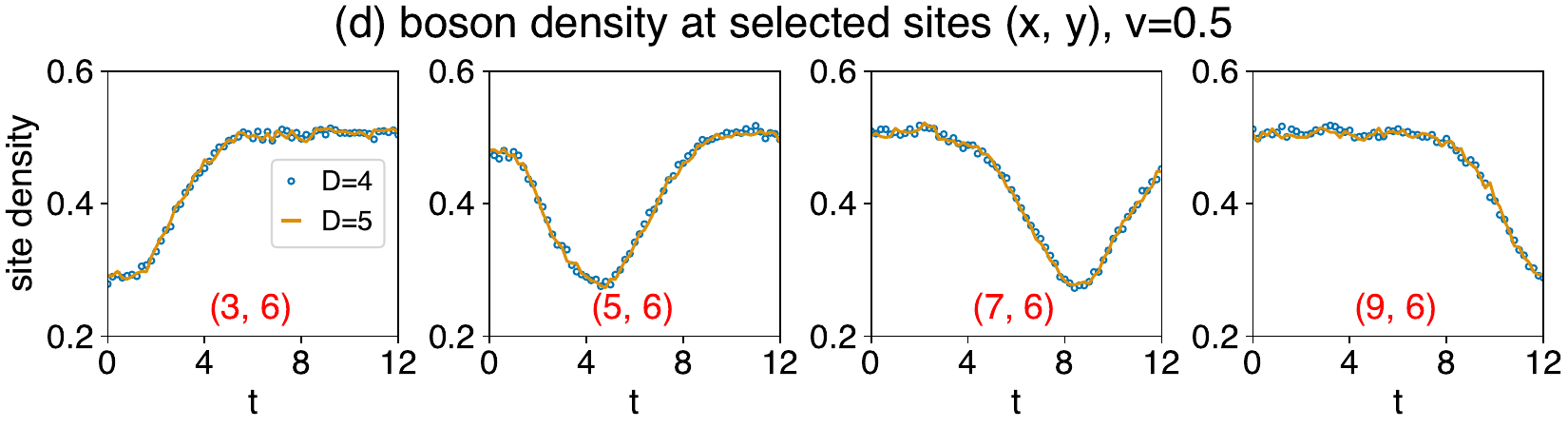}
    \includegraphics[width=\linewidth]{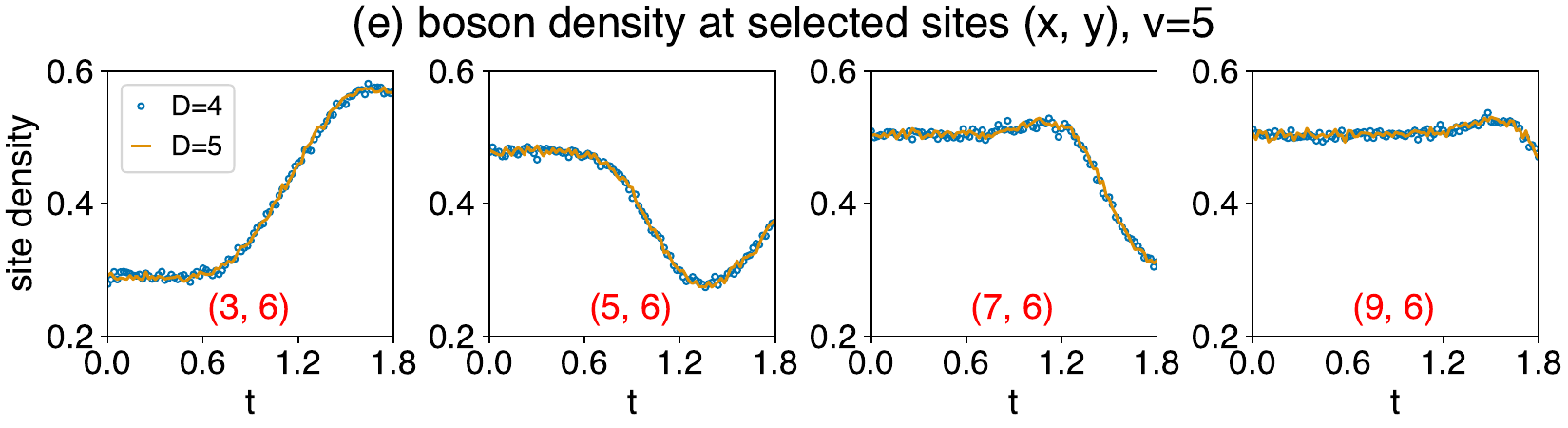}
    \caption{
    (a) Real-time density response of a 2D bosonic superfluid to a moving obstacle (red dot) with velocity $\abs{\vec v} = 0.5$.
Snapshots show the deviation of the local density from 0.5 on a $13\times 13$ lattice for different times $T$ as the obstacle is dragged along the $x$ direction. 
(b) Same as (a), but with $|\vec v|=5$.
(c) Total energy $\braket{H_\text{sf} + H_\text{obs}(t)}$ as a function of time for both velocities. 
To assess the effect of the time step on the energy evolution, results obtained with $dt=0.01$ and $0.02$ are shown. 
For $|\mathbf v|=0.5$, the two time steps agree to within a relative energy difference of $2\times10^{-4}$ over the entire evolution.
(d, e) Superfluid densities at various bond dimensions.
}
    \label{fig:sf}
\end{figure}

Here we study the dynamics of the bosons in the presence of an obstacle, interacting via a repulsive potential: \begin{equation}
  H_\text{obs}(t) = 2 \sum_i b_i^\dag b_i \exp(-\frac{\abs{\vec{r}_i - \vec R(t)}^2}{2})
\end{equation}
where $\vec R(t)$ is the position of the obstacle at time $t$.
By Galilean invariance, we keep the superfluid at rest and drag the obstacle rightward at constant velocity: $\vec R(t) = \vec R_0 + \vec v t$. 
$H_\text{obs}(t)$ can also be realized via a laser in cold-atom simulations.
The initial state is the ground state of $H_\text{sf} + H_\text{obs}(0)$. 
As shown in Fig. \ref{fig:sf} (a,b), for slow motion $|\vec v|=0.5$, the density perturbation remains localized and approximately symmetric around the impurity, with no pronounced wake, consistent with dissipationless flow below the critical velocity. 
For fast motion $|\vec v|=5.0$, the impurity leaves behind a pronounced density wake and radiates excitations throughout the system above the critical velocity.
In particular, the density depletion caused by the obstacle closely follows the obstacle when $\vec v$ is small, while it lags far behind when $\vec v$ is large. 
The superfluid densities are converged with $D=4$ and $5$ as shown in Fig. \ref{fig:sf} (d, e).

Figure~\ref{fig:sf} (c) shows the time evolution of $\braket{H_\text{sf} + H_\text{obs}(t)}$. 
At $\abs{\vec v} = 5$, significant work is done to the boson–obstacle system, leading to pronounced dissipation.
In contrast, for a slow velocity, $\abs{\vec v} = 0.5$, the total energy remains approximately conserved over an extended time interval, consistent with frictionless superfluid motion.
These qualitative differences between the two velocities signal the presence of a critical superfluid velocity, as anticipated by Landau \cite{Landau1941English}.

In the future, it would be exciting to apply PEPS–tVMC for the dynamical studies of phase coherence, vortex nucleation, and other hallmarks of superfluidity and superconductivity.

\subsection{Quantum Ising global quench}
\label{subsec:ising}
Finally we study the dynamics in the quantum Ising model following a global transverse field quench: $h = \infty \rightarrow h_c$ in 
\begin{equation}
  H_\text{ising} = -\sum_{\braket{ij}} \sigma^z_i \sigma^z_j - h \sum_i \sigma^x_i  
\end{equation}
While global quenches are not the main application target of PEPS-tVMC, this Ising dynamics \cite{czarnik2019time} has become a paradigmatic example in the simulation of two-dimensional real-time dynamics, which makes it interesting for PEPS-tVMC to compare. 
On a $11\times11$ lattice, we prepare the initial state as the product state polarized along positive $x$-axis, and evolve it under the critical $H_\text{ising}$ with $h_c=3.04438$ \cite{deng2002}.

\begin{figure}[hbt]
    \centering
    \includegraphics[width=\linewidth]{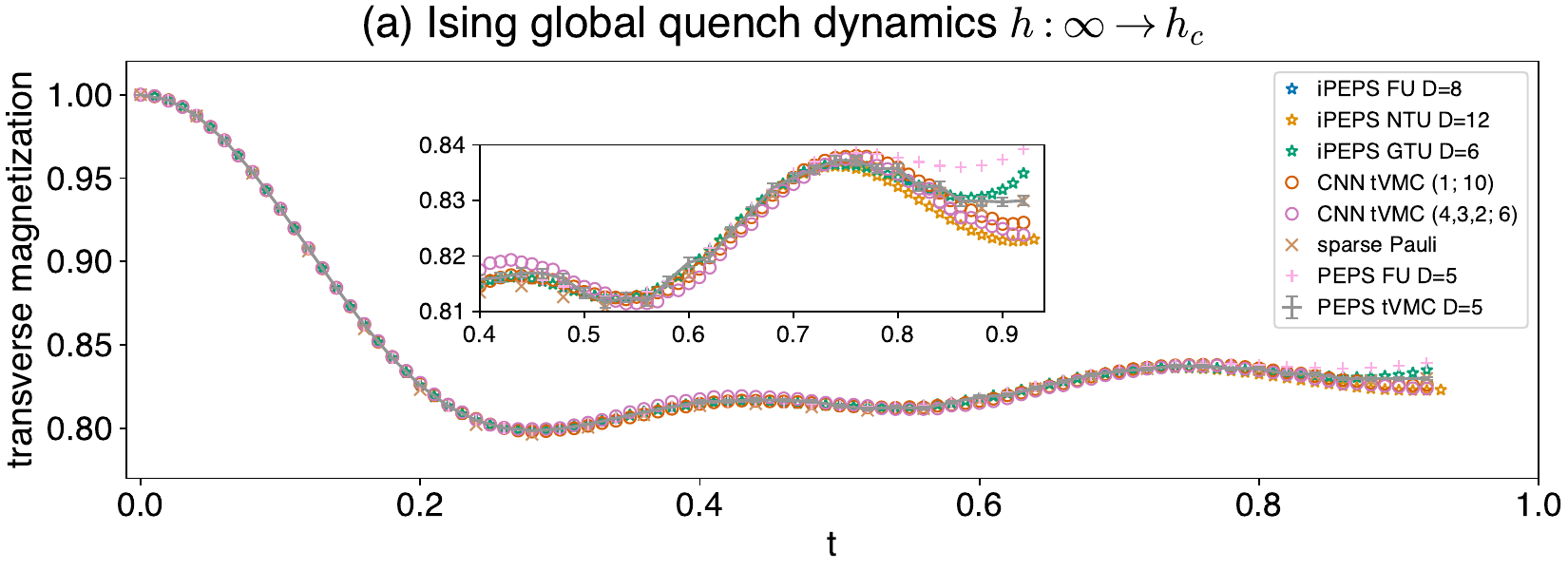}
    \includegraphics[width=\linewidth]{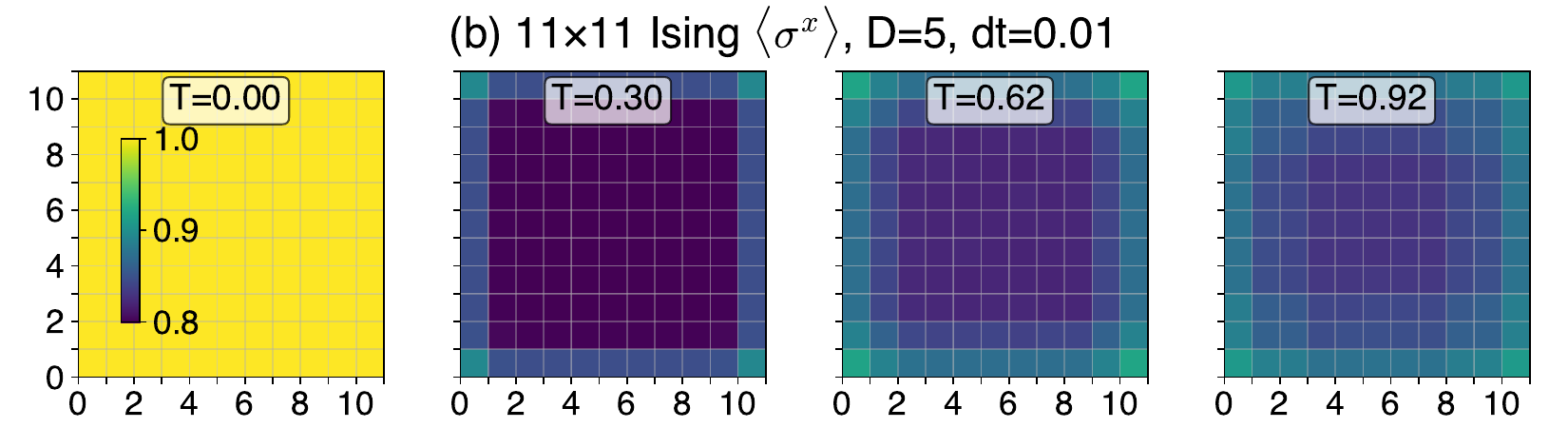}
    \caption{Transverse magnetization during the Ising global quench dynamics.
    The references include iPEPS full update (FU) \cite{czarnik2019time}, neighborhood update (NTU) \cite{dziarmaga2021time}, gradient tensor update (GTU) \cite{dziarmaga2022time}, convolutional neural network (CNN) tVMC \cite{nqstvmc_3}, and sparse Pauli operator dynamics \cite{sparse_pauli}. 
    PEPS FU is done by us at $D=5$ following Ref. \cite{algorithm_finite_peps}.
    }
    \label{fig:ising}
\end{figure}

In Fig. \ref{fig:ising} (a), we report the transverse magnetization $\braket{\sigma^x_{(5,5)}}$ of the middle site during the dynamics.
Fig. \ref{fig:ising} (b) shows the distribution of $\braket{\sigma^x}$ in real-space and suggests that the finite size effect at this time scale is small for the middle site.
This dynamics starts from a product state, which is tricky to represent with a $D=5$ PEPS, so we use FU \cite{full_update_1} to evolve the state first to $T=0.4$, and relay to tVMC to evolve until $T=0.92$.
As a comparison, we also evolve the FU fully to $T=0.92$.  
We use bond dimension 125 for the bMPO of the FU, and 20 for the bMPS of the tVMC.
Comparing with previous results, tVMC is likely to be the more accurate of the two.

At the final time $T = 0.92$, various simulations \cite{czarnik2019time,dziarmaga2021time,dziarmaga2022time,nqstvmc_3,sparse_pauli}, including ours, show a discrepancy of around 0.005 to 0.01 in $\braket{\sigma^x}$.  
We can in fact evolve the PEPS--tVMC trajectory past $T=0.92$; the results are not shown because, beyond this time, the available methods begin to disagree and it is no longer clear which calculation should be taken as the benchmark. 
It will remain a collective effort of the community to close out this discrepancy in the future.

Finally, we comment on a practical strategy to scale up the PEPS-tVMC to high bond dimensions for this quench.
Currently all PEPS tensors are allowed to vary independently, amounting to a large number of parameters: $N_p = 2L^2 D^4$, incurring a large memory cost if the sample size is also large. 
It is possible to variationally constrain the tensors within the boundary to be the same, reducing $N_p$ to order $O(D^4)$. 
In the presence of an open boundary, this will introduce a bias. 
However, due to locality, one expects that the observables deep in the bulk to be unaffected by such bias before a certain time-scale and system size.
This would possibly allow the PEPS-tVMC to be scaled up to significantly higher bond dimension and system sizes, which we leave for future work.
Such higher-bond-dimension calculations should also help resolve the discrepancies among different methods beyond the present time window of agreement.

\subsection{Error and stability in solving the TDVP equation}
An important factor underlying the accuracy and efficiency of PEPS--tVMC is the numerical stability with which the TDVP equation---or its reduced minSR variant---can be solved.
All simulations in this work solve the TDVP equation using a Cholesky decomposition \cite{golub2013matrix} of the QGT matrix with a very small Tikhonov regulator ($10^{-7}$ or $10^{-8}$).

Compared to the rank-revealing singular value decomposition (SVD) or pseudo-inverse solver, the Cholesky solver lacks the robustness of SVD  or pseudo-inverse when applied to ill-conditioned matrices, but is substantially more parallelizable and faster on GPUs.

For example, on a GPU card with 141 GB memory and 34 TFLOPS for FP64, for a $20480\times20480$ positive $\texttt{complex128}$ matrix $S$, solving the linear equation $S \vec x = \vec g$ takes 146 seconds  with the SVD solver, but only 0.28 seconds with the Cholesky solver.
At $40960 \times 40960$, the Cholesky solver takes 2.1 seconds, while the SVD solver throws an out-of-memory error.
Such equations need to be solved $k$ times in a $k$th order RK integrator. 
See repository \cite{data} for the benchmark details.

Recent NQS--tVMC studies often resort to the SVD solver~\cite{nqstvmc_3,nqstvmc_7,Lange_2024}, trading computational efficiency for numerical robustness.
The fact that Cholesky remains stable in all our simulations indicates that once gauge redundancies are analytically removed, the PEPS variational manifold becomes exceptionally stable for real-time dynamics, which gives PEPS-tVMC a significant walltime reduction.

\begin{figure}[hbt]
    \centering
    \includegraphics[width=\linewidth]{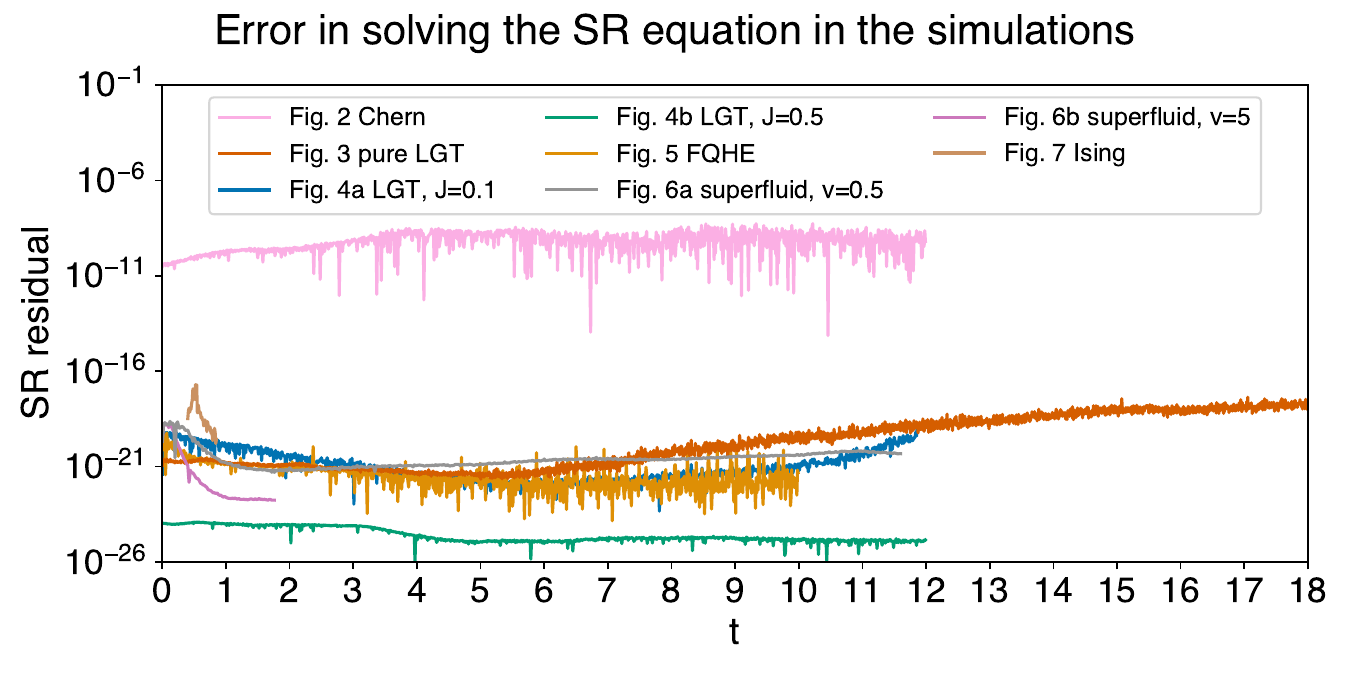}
    \caption{
    Residual error $\varepsilon = \| S\dot{\boldsymbol{\theta}} - i\mathbf{g} \|^{2} / \|\mathbf{g}\|^{2}$ during PEPS--tVMC time evolution for the models considered in this work.}
    \label{fig:sr_error}
\end{figure}

To quantify the accuracy of the Cholesky solution, we monitor the tVMC residual with the regulator removed,
\begin{equation}
  \varepsilon = \frac{\| S\dot{\vec\theta} - (-i\vec g) \|^{2}}{\| \vec g \|^{2}},
\end{equation}
which measures the relative error in satisfying the TDVP equation.
As shown in Fig.~\ref{fig:sr_error}, the residual remains exceptionally small---typically between $10^{-9}$ and $10^{-25}$, depending on the model.
These results confirm the effectiveness of our robust strategies to solve the linear TDVP equations, ensuring that the tVMC step does not contribute appreciable error during real-time evolution.

\subsection{Heuristic reason for the stability}
We give a preliminary understanding of why PEPS-tVMC is stable related to the small-$o$ trick. 
Suppose we are in the minSR regime, then the stability of solving the tVMC equation is determined by the condition number of $OO^\dag$, which is a measure how ``orthogonal'' the row vectors of $O$ are to each other.  
For two row vectors corresponding to sample $\vec s$ and $\vec s'$, due to Eq. \ref{eq:smallo_0}, their overlap only receives contributions from the sites $\vec x$ for which $\vec s(\vec x) = \vec s'(\vec x)$: 
\begin{equation*}
  \sum_{\alpha} O_\alpha^*(\vec s) O_\alpha(\vec s') = \sum_{\vec x: \, \vec s(\vec x) 
  = \vec s'(\vec x)} o^\dag[\vec x](\vec s) o[\vec x](\vec s')
\end{equation*}
Roughly speaking, out of the $\sim L^2D^4d$ parameters of a PEPS, only $\frac{L^2}{d}D^4$ number of them contribute to the pair-wise overlap of the row vectors of $O$. 

In general, if each sample only activates $\frac{1}{d}$ of total parameters of an ansatz, then roughly $\frac{1}{d^2}$ of total parameters will contribute to the off-diagonal elements of $OO^\dag$, while $\frac{1}{d}$ of them to the diagonal.

Amusingly, the above mechanism is reminiscent of mixture-of-experts (MoE) models \cite{jacobs1991adaptive,jordan1994hierarchical,shazeer2017outrageously,fedus2022switch}.
In a MoE model, the input $\vec s$, via a rounter, selects a subset of experts $E$, the output is a weighted sum of the selected expert outputs,
\begin{equation}
  \mathrm{MoE}(\vec s)=\sum_{a \in E} g_a(\vec s) E_a(\vec s).
\end{equation}
In the small-$o$ PEPS representation, on each site, one PEPS tensor may be viewed as $d$ local experts with disjoint parameters. 
A sample $\vec s$ selects exactly one local expert, and instead of a weighted sum, the final output is the single layer contraction of the selected $L^2$ experts out of a total of $dL^2$ experts,
\begin{equation}
  \Psi(\vec s)= \text{tTr} \left(\prod_{\vec x} o[\vec x](\vec s)\right)
\end{equation}

\subsection{Simulation parameters}
We document the simulation details in Table \ref{tab:parameters}. 
The integrator is 2nd order RK for pure $\Z_2$ gauge theory and 4th order RK for all other examples.
The Tikhonov regulator in solving the TDVP or the minSR equation using the Cholesky decomposition is $10^{-8}$ everywhere, except for the $D=4$ pure $\Z_2$ gauge theory which is $10^{-7}$.

For the $\Z_2$ lattice gauge theory, despite the large $N_p$ due to $D=10$, the computational complexity is at the level of a $D=5$ normal PEPS.
The time saving is from our findings in Ref. \cite{wu2025accurate}, while the memory saving is due to the small-$o$ trick.
If one stores the full $20480 \times 1027320$ \texttt{complex128} $O$ matrix, the memory cost would be 336GB, which is a significant burden.

\begin{table}[!t]
\renewcommand{\arraystretch}{1.2}
\begin{tabular}{lcccccccc}
\hline\hline
Model
& $L$ 
& $D$
& $D'$
& $\mathrm{d}t$
& $T$
& $N_s$
& $N_p$
& $t_{\rm wall}$/s \\
\hline

Chern
& 12 & 4 & 16 & 1e-2 & 12 & 40960 & 28k & $399\times1$ \\

pure LGT
& 10 & 4 & 8  & 5e-3 & 10 & 40960 & 8k   & $17\times1$ \\
&    & 6 & 9  &       &    & 10240 & 42k  & $13\times1$ \\
&    & 8 & 12 &       &    &       & 134k & $38\times1$ \\

Higgs LGT
& 12 & 6  & 12 & 1e-2 & 12 & 20480 & 134k  & $167\times1$ \\
&    & 8  & 16 &      &    &       & 422k  & $328\times1$ \\
&    & 10 & 20 &      &    &       & 1027k & $560\times2$ \\

Hofstadter
& 12 & 4 & 24 & 1e-2 & 10 & 20480 & 52k  & $207\times2$ \\
&    & 5 & 30 &      &    &       & 129k & $538\times4$ \\

superfluid
& 13 & 4 & 16 & 2e-2 & 12 & 20480 & 63k  & $222\times1$ \\
&    & 5 & 20 &      &    &       & 154k & $713\times1$ \\

Ising
& 11 & 5 & 20 & 1e-2 & 0.92 & 40960 & 105k &  $430\times2$ \\
\hline\hline
\end{tabular}
\caption{
Simulation parameters.
$L$: side length.
$D$: PEPS bond dimension.
$D'$: bMPS bond dimension.
$\mathrm{d}t$: time step.
$T$: evolution time.
$N_s$: sample size.
$N_p$: parameter size in thousands.
$t_{\rm wall}$: wall time per $\mathrm{d}t$ in seconds; $\times n$ means the simulation ran on $n$ GPU cards.
The wall time is measure on GPUs with 141 GB memory and 34 TFLOPS for FP64.
We use the standard TDVP if $N_s > N_p$, and minSR otherwise.
Identical parameters for the same model are not shown.
}
\label{tab:parameters}
\end{table}

\section{Conclusion}
\label{sec:discussion}
In this work we have shown that projected entangled pair states, when combined with time-dependent variational Monte Carlo, provide a stable and efficient framework for simulating real-time low-energy dynamics in two-dimensional quantum many-body systems.
The key technical insight is that the ill-conditioning traditionally associated with PEPS---stemming from gauge redundancy---can be removed analytically in closed form.
Once gauge directions are eliminated, the TDVP equation can be solved efficiently with high numerical stability over long evolution times.
This stability, together with the locality and single-layer contraction structure of PEPS within VMC, makes PEPS--tVMC a powerful classical method for studying low-energy dynamics in 2D.

The examples explored in Sec.~\ref{sec:results} demonstrate the breadth of
physical phenomena accessible with this approach.
These results establish PEPS--tVMC as a general-purpose tool for simulating elementary excitations, transport phenomena, and real-time response in a wide range of quantum systems with realistic computational resources.
The improvements developed here apply equally to ground-state PEPS--VMC calculations \cite{liu2022gapless,liu2022emergence,liu2024quantum,liu2025hubbard}.

An important conceptual question concerns the regime of validity of PEPS for
real-time dynamics.
While PEPS cannot represent volume-law states, our results indicate that for
local quenches and low-energy excitations the entanglement growth remains
compatible with PEPS bond dimensions accessible on current hardware.
A more systematic understanding of entanglement-light-cone structure in 2D
local quenches, and of the interplay between PEPS bond dimension and dynamical
fidelity, would provide valuable theoretical guidance.
In addition, studying late-time thermalization or hydrodynamic transport within
PEPS--tVMC may reveal whether approximate area-law manifolds can capture
emergent long-wavelength physics even when the exact state exhibits extensive
entanglement.

Overall, our work demonstrates that PEPS, long regarded as difficult for
time evolution, can in fact support accurate and stable simulations of
large-scale 2D quantum dynamics at low energy.
We hope that the ideas introduced here can inspire new developments in both variational algorithms and tensor-network theory, ultimately expanding the reach of classical simulation deep into the low-energy real-time dynamics of quantum matter.

\section{Open question and future direction}
\label{sec:open}
We close by listing a few concrete open questions and future directions regarding the broad scope of PEPS-tVMC.

Open questions:
\begin{enumerate}
  \item \textit{Support collapse.} tVMC has a known problem of ``support collapse'' in evolving a state acted by a projective operator \cite{nqstvmc_6}.
    For example, the action of $c^\dag_i$ annihilates the sampling probability of any configuration with $s_i = 0$, which introduces a bias in the stochastic estimation of the QGT. 
\footnote{At the time of the writing, we have made considerable progress in solving this problem. We expect an elegant solution of the local support collapse problem for PEPS-tVMC.}
  \item \textit{Product state.} Unlike SU or FU, tVMC cannot grow bond dimension dynamically.
  If an evolution starts from a product state, one cannot simply use a $D=1$ PEPS due to the entanglement later in the dynamics. 
  The naive zero-padding of a product state into a higher $D$ PEPS will not work as the dynamics will be trapped in the local stationary points of the TDVP.
  We mention two possible strategies: i) one may use the standard PEPS time evolution algorithm such as SU or FU for early time with a small bMPO and switch to tVMC later when the required bMPO becomes too expensive, as already used in Fig. \ref{fig:ising}; ii) one may prepare the product state by a ground state VMC of its parent Hamiltonian made of one-site terms.
\item \textit{Amplitude evaluation.} PEPS-VMC hinges on an accurate estimate of the amplitude $\braket{\vec s|\Psi}$. 
  While the bMPS method already does quite well, can one do better?
\item \textit{PEPS versus NQS.} 
  What features of PEPS and NQS control the conditioning of the tVMC equations?
  How is this related to their different entanglement structures, including the ability of NQS to represent volume-law states?
  At the same time, PEPS offer several practical advantages as a tVMC ansatz: the accuracy can be systematically improved by increasing the bond dimension, and the method connects naturally to established PEPS time-evolution techniques, including iPEPS updates, finite-PEPS full update and belief propagation.
\end{enumerate}

Future directions:
\begin{enumerate}
  \item \textit{Long-range interactions.}
  Long-range interacting systems, such as Rydberg-atom arrays, have recently attracted considerable attention \cite{browaeys2020manybody}.
  Incorporating long-range interactions into tVMC is conceptually straightforward, since the action of long-range couplings only enters through local energy estimators.
  While new ideas would be desirable to address the increased computational cost from long-range off-diagonal couplings, models with local off-diagonal terms and long-range diagonal couplings are already within the reach of the method, as considered in Ref. \cite{mauron2025predicting}.
  \item \textit{Periodic boundary conditions.}
  Extending PEPS--tVMC to periodic boundary conditions is possible in principle, but the cost of amplitude evaluation of a boundaryless PEPS single layer is substantially higher.
  One may find inspirations in the large body of work in the tensor network renormalization of two-dimensional classical partition functions \cite{levin2007tensor,gu2009tensor,evenbly2015tensor,zhao2010renormalization}, or the more recent developments of belief propagation \cite{alkabetz2021belief,guo2023blockbp,evenbly2026loop}.
  \item \textit{Finite-temperature calculation.}
  PEPS--tVMC should be useful for imaginary-time dynamics as well.
  Two possible routes are the minimally entangled typical thermal state approach \cite{white2009metts,stoudenmire2010metts} and thermofield-double purification \cite{feiguin2005finite,zhang2025scalable}.
  \item \textit{Adiabatic dynamics.}
  PEPS--tVMC should also be well suited for adiabatic dynamics, which is particularly relevant to how current quantum computers and simulators prepare many-body ground states \cite{albash2018adiabatic}.
  In this setting, the system follows (near) ground states of a time-dependent Hamiltonian, which are expected to obey an area law.
  \item \textit{Spectral functions.}
  Single-particle spectral functions and density response functions are directly connected to experiments.
  Once local support collapse is resolved, PEPS--tVMC could become a practical dynamical probe of two-dimensional strongly correlated systems.
  \item \textit{Single precision.}
  The simulations in this work use double precision.
  However, modern GPUs provide much higher throughput for single precision.
  It is therefore important to push, as much as possible, PEPS--tVMC to be performed in lower precision.
\end{enumerate}

\begin{acknowledgements}
Part of the algorithm is tested using the TenPy code base \cite{hauschild2018efficient, johannes2024tensor}.
The code for production is implemented with JAX \cite{jax2018github}.
Y.W. is supported by a startup grant from the IOP-CAS.
Y.W. thanks his colleagues Yu-Xiang Zhang, Shi-Xin Zhang, and Tao Xiang for sharing computational resources with him.
He benefited greatly from discussion and collaboration on various related projects with Bartholomew Andrews, Zhehao Dai, Sheng-Hsuan Lin, Sajant Anand, Taige Wang, Michael Zaletel, Frank Pollmann, Wen-Yuan Liu, Jiahang Hu, Jiajun Yu, and Yuntian Gu. J.N. thanks Juan Carrasquilla for insightful discussions.
After the paper appeared as a preprint, we benefited from questions, comments, and discussion from Giuseppe Carleo, Markus Schmitt, Miles Stoudenmire, Joseph Tindall, Jad Halimeh, Qi Yang, and Yijia Wang.   
We thank Jacek Dziarmaga and Markus Schmitt for providing the Ising dynamics data of iPEPS and CNN-tVMC.
We thank Sheng-Hsuan Lin for an earlier version of the PEPS FU code which we adopted for the Ising dynamics; we alone are responsible for any potential mistakes of the FU simulation.
The data that support the findings of this article are openly available \cite{data}; embargo periods may apply.
The simulations were supported by the IOP-CAS computing facilities.
\end{acknowledgements}
\bibliography{ref.bib}

@article{zhao2010renormalization,
  title = {Renormalization of tensor-network states},
  author = {Zhao, H. H. and Xie, Z. Y. and Chen, Q. N. and Wei, Z. C. and Cai, J. W. and Xiang, T.},
  journal = {Phys. Rev. B},
  volume = {81},
  issue = {17},
  pages = {174411},
  numpages = {17},
  year = {2010},
  month = {May},
  publisher = {American Physical Society},
  doi = {10.1103/PhysRevB.81.174411},
  url = {https://link.aps.org/doi/10.1103/PhysRevB.81.174411}
}

@article{Calabrese2005Evolution,
  author  = {Calabrese, Pasquale and Cardy, John},
  title   = {Evolution of entanglement entropy in one-dimensional systems},
  journal = {Journal of Statistical Mechanics: Theory and Experiment},
  volume  = {2005},
  number  = {04},
  pages   = {P04010},
  year    = {2005},
  doi     = {10.1088/1742-5468/2005/04/P04010},
  eprint  = {cond-mat/0503393},
  archivePrefix = {arXiv}
}

@article{tindall2024efficient,
  title = {Efficient Tensor Network Simulation of {IBM}'s Eagle Kicked Ising Experiment},
  author = {Tindall, Joseph and Fishman, Matt and Stoudenmire, Miles and Sels, Dries},
  journal = {PRX Quantum},
  volume = {5},
  issue = {1},
  pages = {010308},
  year = {2024},
  doi = {10.1103/PRXQuantum.5.010308}
}

@article{begusic2024fast,
  title = {Fast and converged classical simulations of evidence for the utility of quantum computing before fault tolerance},
  author = {Begu{\v{s}}i{\'c}, Tomislav and Gray, Johnnie and Chan, Garnet Kin-Lic},
  journal = {Science Advances},
  volume = {10},
  number = {3},
  pages = {eadk4321},
  year = {2024},
  doi = {10.1126/sciadv.adk4321}
}

@article{tindall2025dynamics,
  title = {Dynamics of disordered quantum systems with two- and three-dimensional tensor networks},
  author = {Tindall, Joseph and Mello, Antonio and Fishman, Matt and Stoudenmire, Miles and Sels, Dries},
  journal = {arXiv preprint arXiv:2503.05693},
  year = {2025},
  eprint = {2503.05693},
  archivePrefix = {arXiv},
  primaryClass = {quant-ph}
}

@article{lee2025scalable,
  title = {Scalable projected entangled-pair state representation of random quantum circuit states},
  author = {Lee, Sung-Bin B. and Choi, Hee Ryang and Ohm, Daniel Donghyon and Lee, Seung-Sup B.},
  journal = {Physical Review Research},
  volume = {7},
  number = {3},
  pages = {033252},
  year = {2025},
  doi = {10.1103/rzgm-cywf}
}

@article{luchnikov2024largescale,
  title = {Large-scale quantum annealing simulation with tensor networks and belief propagation},
  author = {Luchnikov, Ilia A. and Tiunov, Egor S. and Haug, Tobias and Aolita, Leandro},
  journal = {arXiv preprint arXiv:2409.12240},
  year = {2024},
  eprint = {2409.12240},
  archivePrefix = {arXiv},
  primaryClass = {quant-ph}
}

@article{mauron2025predicting,
  title = {Predicting topological entanglement entropy in a {R}ydberg analogue simulator},
  author = {Mauron, Linda and Denis, Zakari and Nys, Jannes and Carleo, Giuseppe},
  journal = {Nature Physics},
  volume = {21},
  pages = {1332--1337},
  year = {2025},
  doi = {10.1038/s41567-025-02944-3}
}

@article{browaeys2020manybody,
  title = {Many-body physics with individually controlled {R}ydberg atoms},
  author = {Browaeys, Antoine and Lahaye, Thierry},
  journal = {Nature Physics},
  volume = {16},
  pages = {132--142},
  year = {2020},
  doi = {10.1038/s41567-019-0733-z}
}

@article{levin2007tensor,
  title = {Tensor renormalization group approach to two-dimensional classical lattice models},
  author = {Levin, Michael and Nave, Cody P.},
  journal = {Physical Review Letters},
  volume = {99},
  pages = {120601},
  year = {2007},
  doi = {10.1103/PhysRevLett.99.120601}
}

@article{gu2009tensor,
  title = {Tensor-entanglement-filtering renormalization approach and symmetry-protected topological order},
  author = {Gu, Zheng-Cheng and Wen, Xiao-Gang},
  journal = {Physical Review B},
  volume = {80},
  pages = {155131},
  year = {2009},
  doi = {10.1103/PhysRevB.80.155131}
}

@article{evenbly2015tensor,
  title = {Tensor network renormalization},
  author = {Evenbly, Glen and Vidal, Guifr{\'e}},
  journal = {Physical Review Letters},
  volume = {115},
  pages = {180405},
  year = {2015},
  doi = {10.1103/PhysRevLett.115.180405}
}

@article{alkabetz2021belief,
  title = {Tensor networks contraction and the belief propagation algorithm},
  author = {Alkabetz, Roy and Arad, Itai},
  journal = {Physical Review Research},
  volume = {3},
  pages = {023073},
  year = {2021},
  doi = {10.1103/PhysRevResearch.3.023073}
}

@article{guo2023blockbp,
  title = {Block belief propagation algorithm for two-dimensional tensor networks},
  author = {Guo, Chu and Poletti, Dario and Arad, Itai},
  journal = {Phys. Rev. B},
  volume = {108},
  issue = {12},
  pages = {125111},
  numpages = {12},
  year = {2023},
  month = {Sep},
  publisher = {American Physical Society},
  doi = {10.1103/PhysRevB.108.125111},
  url = {https://link.aps.org/doi/10.1103/PhysRevB.108.125111}
}

@article{white2009metts,
  title = {Minimally Entangled Typical Quantum States at Finite Temperature},
  author = {White, Steven R.},
  journal = {Physical Review Letters},
  volume = {102},
  pages = {190601},
  year = {2009},
  doi = {10.1103/PhysRevLett.102.190601}
}

@article{stoudenmire2010metts,
  title = {Minimally entangled typical thermal state algorithms},
  author = {Stoudenmire, E. M. and White, Steven R.},
  journal = {New Journal of Physics},
  volume = {12},
  pages = {055026},
  year = {2010},
  doi = {10.1088/1367-2630/12/5/055026}
}

@article{feiguin2005finite,
  title = {Finite-temperature density matrix renormalization using an enlarged Hilbert space},
  author = {Feiguin, Adrian E. and White, Steven R.},
  journal = {Physical Review B},
  volume = {72},
  pages = {220401},
  year = {2005},
  doi = {10.1103/PhysRevB.72.220401}
}

@article{albash2018adiabatic,
  title = {Adiabatic quantum computation},
  author = {Albash, Tameem and Lidar, Daniel A.},
  journal = {Reviews of Modern Physics},
  volume = {90},
  pages = {015002},
  year = {2018},
  doi = {10.1103/RevModPhys.90.015002}
}

@article{jacobs1991adaptive,
  title   = {Adaptive Mixtures of Local Experts},
  author  = {Jacobs, Robert A. and Jordan, Michael I. and Nowlan, Steven J. and Hinton, Geoffrey E.},
  journal = {Neural Computation},
  volume  = {3},
  number  = {1},
  pages   = {79--87},
  year    = {1991},
  doi     = {10.1162/neco.1991.3.1.79}
}

@article{jordan1994hierarchical,
  title   = {Hierarchical Mixtures of Experts and the {EM} Algorithm},
  author  = {Jordan, Michael I. and Jacobs, Robert A.},
  journal = {Neural Computation},
  volume  = {6},
  number  = {2},
  pages   = {181--214},
  year    = {1994},
  doi     = {10.1162/neco.1994.6.2.181}
}

@inproceedings{shazeer2017outrageously,
  title     = {Outrageously Large Neural Networks: The Sparsely-Gated Mixture-of-Experts Layer},
  author    = {Shazeer, Noam and Mirhoseini, Azalia and Maziarz, Krzysztof and Davis, Andy and Le, Quoc V. and Hinton, Geoffrey E. and Dean, Jeff},
  booktitle = {International Conference on Learning Representations},
  year      = {2017},
  url       = {https://openreview.net/forum?id=B1ckMDqlg}
}

@article{fedus2022switch,
  title   = {Switch Transformers: Scaling to Trillion Parameter Models with Simple and Efficient Sparsity},
  author  = {Fedus, William and Zoph, Barret and Shazeer, Noam},
  journal = {Journal of Machine Learning Research},
  volume  = {23},
  number  = {120},
  pages   = {1--39},
  year    = {2022}
}

@article{vison,
  title = {${Z}_{2}$ gauge theory of electron fractionalization in strongly correlated systems},
  author = {Senthil, T. and Fisher, Matthew P. A.},
  journal = {Phys. Rev. B},
  volume = {62},
  issue = {12},
  pages = {7850--7881},
  numpages = {0},
  year = {2000},
  month = {Sep},
  publisher = {American Physical Society},
  doi = {10.1103/PhysRevB.62.7850},
  url = {https://link.aps.org/doi/10.1103/PhysRevB.62.7850}
}

@article{deng2002,
  title = {Cluster Monte Carlo simulation of the transverse Ising model},
  author = {Bl\"ote, Henk W. J. and Deng, Youjin},
  journal = {Phys. Rev. E},
  volume = {66},
  issue = {6},
  pages = {066110},
  numpages = {8}, 
  year = {2002}, 
  month = {Dec},
  publisher = {American Physical Society},
  doi = {10.1103/PhysRevE.66.066110},
  url = {https://link.aps.org/doi/10.1103/PhysRevE.66.066110}
}

@article{sparse_pauli,
  title = {Real-Time Operator Evolution in Two and Three Dimensions via Sparse Pauli Dynamics},
  author = {Begu\ifmmode \check{s}\else \v{s}\fi{}i\ifmmode \acute{c}\else \'{c}\fi{}, Tomislav and Chan, Garnet Kin-Lic},
  journal = {PRX Quantum},
  volume = {6},
  issue = {2},
  pages = {020302},
  numpages = {13},
  year = {2025},
  month = {Apr},
  publisher = {American Physical Society},
  doi = {10.1103/PRXQuantum.6.020302},
  url = {https://link.aps.org/doi/10.1103/PRXQuantum.6.020302}
}

@misc{garratt2026quantum,
      title={Quantum matter is weakly entangled at low energies}, 
      author={Samuel J. Garratt and Dmitry A. Abanin},
      year={2026},
      eprint={2604.14143},
      archivePrefix={arXiv},
      primaryClass={cond-mat.stat-mech},
      url={https://arxiv.org/abs/2604.14143}, 
}

@misc{grover2026hierarchical,
      title={Hierarchical entanglement transitions and hidden area-law sectors in quantum many-body dynamics},
      author={Tarun Grover},
      year={2026},
      eprint={2605.04540},
      archivePrefix={arXiv},
      primaryClass={quant-ph},
      url={https://arxiv.org/abs/2605.04540},
}

@Article{johannes2024tensor,
	title={{Tensor network Python (TeNPy) version 1}},
	author={Johannes Hauschild and Jakob Unfried and Sajant Anand and Bartholomew Andrews and Marcus Bintz and Umberto Borla and Stefan Divic and Markus Drescher and Jan Geiger and Martin Hefel and Kévin Hémery and Wilhelm Kadow and Jack Kemp and Nico Kirchner and Vincent S. Liu and Gunnar Möller and Daniel Parker and Michael Rader and Anton Romen and Samuel Scalet and Leon Schoonderwoerd and Maximilian Schulz and Tomohiro Soejima and Philipp Thoma and Yantao Wu and Philip Zechmann and Ludwig Zweng and Roger S. K. Mong and Michael P. Zaletel and Frank Pollmann},
	journal={SciPost Phys. Codebases},
	pages={41},
	year={2024},
	publisher={SciPost},
	doi={10.21468/SciPostPhysCodeb.41},
	url={https://scipost.org/10.21468/SciPostPhysCodeb.41},
}

@software{jax2018github,
  author = {James Bradbury and Roy Frostig and Peter Hawkins and Matthew James Johnson and Chris Leary and Dougal Maclaurin and George Necula and Adam Paszke and Jake Vander{P}las and Skye Wanderman-{M}ilne and Qiao Zhang},
  title = {{JAX}: composable transformations of {P}ython+{N}um{P}y programs},
  url = {http://github.com/jax-ml/jax},
  version = {0.3.13},
  year = {2018},
}

@incollection{Sandvik2010QMC,
  title        = {Computational Studies of Quantum Spin Systems},
  author       = {Sandvik, Anders W.},
  booktitle    = {AIP Conference Proceedings},
  volume       = {1297},
  pages        = {135--338},
  year         = {2010},
  doi          = {10.1063/1.3518900}
}

@book{BeccaSorella2017QMCBook,
  title     = {Quantum Monte Carlo Approaches for Correlated Systems},
  author    = {Becca, Federico and Sorella, Sandro},
  publisher = {Cambridge University Press},
  year      = {2017},
  doi       = {10.1017/9781316417041}
}

@article{SR1998,
  title = {Green Function Monte Carlo with Stochastic Reconfiguration},
  author = {Sorella, Sandro},
  journal = {Phys. Rev. Lett.},
  volume = {80},
  issue = {20},
  pages = {4558--4561},
  numpages = {0},
  year = {1998},
  month = {May},
  publisher = {American Physical Society},
  doi = {10.1103/PhysRevLett.80.4558},
  url = {https://link.aps.org/doi/10.1103/PhysRevLett.80.4558}
}

@Article{QGT,
author={Provost, J. P.
and Vallee, G.},
title={Riemannian structure on manifolds of quantum states},
journal={Communications in Mathematical Physics},
year={1980},
month={Sep},
day={01},
volume={76},
number={3},
pages={289-301},
issn={1432-0916},
doi={10.1007/BF02193559},
url={https://doi.org/10.1007/BF02193559}
}

@book{sorella2017,
  title={Quantum Monte Carlo Approaches for Correlated Systems},
  author={ Becca, Federico and Sorella,Sandro },
  year={2017},
  publisher={Cambridge University Press},
  city={ambridge}
}

@article{sorella2001generalized,
  title = {Generalized Lanczos algorithm for variational quantum Monte Carlo},
  author = {Sorella, Sandro},
  journal = {Phys. Rev. B},
  volume = {64},
  issue = {2},
  pages = {024512},
  numpages = {16},
  year = {2001},
  month = {Jun},
  publisher = {American Physical Society},
  doi = {10.1103/PhysRevB.64.024512},
  url = {https://link.aps.org/doi/10.1103/PhysRevB.64.024512}
}

@Article{tvmc1,
author={Carleo, Giuseppe
and Becca, Federico
and Schir{\'o}, Marco
and Fabrizio, Michele},
title={Localization and Glassy Dynamics Of Many-Body Quantum Systems},
journal={Scientific Reports},
year={2012},
month={Feb},
day={06},
volume={2},
number={1},
pages={243},
issn={2045-2322},
doi={10.1038/srep00243},
url={https://doi.org/10.1038/srep00243}
}

@article{tvmc2,
  title = {Light-cone effect and supersonic correlations in one- and two-dimensional bosonic superfluids},
  author = {Carleo, Giuseppe and Becca, Federico and Sanchez-Palencia, Laurent and Sorella, Sandro and Fabrizio, Michele},
  journal = {Phys. Rev. A},
  volume = {89},
  issue = {3},
  pages = {031602},
  numpages = {5},
  year = {2014},
  month = {Mar},
  publisher = {American Physical Society},
  doi = {10.1103/PhysRevA.89.031602},
  url = {https://link.aps.org/doi/10.1103/PhysRevA.89.031602}
}

@article{tvmc3,
  title = {Time-dependent many-variable variational Monte Carlo method for nonequilibrium strongly correlated electron systems},
  author = {Ido, Kota and Ohgoe, Takahiro and Imada, Masatoshi},
  journal = {Phys. Rev. B},
  volume = {92},
  issue = {24},
  pages = {245106},
  numpages = {11},
  year = {2015},
  month = {Dec},
  publisher = {American Physical Society},
  doi = {10.1103/PhysRevB.92.245106},
  url = {https://link.aps.org/doi/10.1103/PhysRevB.92.245106}
}

@article{tvmc_4,
  title = {Unitary Dynamics of Strongly Interacting Bose Gases with the Time-Dependent Variational Monte Carlo Method in Continuous Space},
  author = {Carleo, Giuseppe and Cevolani, Lorenzo and Sanchez-Palencia, Laurent and Holzmann, Markus},
  journal = {Phys. Rev. X},
  volume = {7},
  issue = {3},
  pages = {031026},
  numpages = {12},
  year = {2017},
  month = {Aug},
  publisher = {American Physical Society},
  doi = {10.1103/PhysRevX.7.031026},
  url = {https://link.aps.org/doi/10.1103/PhysRevX.7.031026}
}

@article{minsr,
  title   = {Empowering deep neural quantum states through efficient optimization},
  author  = {Chen, Ao and Heyl, Markus},
  journal = {Nature Physics},
  volume  = {20},
  pages   = {1476--1481},
  year    = {2024},
  doi     = {10.1038/s41567-024-02566-1}
}

@article{nqs,
author = {Giuseppe Carleo  and Matthias Troyer },
title = {Solving the quantum many-body problem with artificial neural networks},
journal = {Science},
volume = {355},
number = {6325},
pages = {602-606},
year = {2017},
doi = {10.1126/science.aag2302},
}

@article{nqstvmc_3,
  title = {Quantum Many-Body Dynamics in Two Dimensions with Artificial Neural Networks},
  author = {Schmitt, Markus and Heyl, Markus},
  journal = {Phys. Rev. Lett.},
  volume = {125},
  issue = {10},
  pages = {100503},
  numpages = {7},
  year = {2020},
  month = {Sep},
  publisher = {American Physical Society},
  doi = {10.1103/PhysRevLett.125.100503},
  url = {https://link.aps.org/doi/10.1103/PhysRevLett.125.100503}
}

@article{nqstvmc_4,
  doi = {10.22331/q-2022-01-20-627},
  url = {https://doi.org/10.22331/q-2022-01-20-627},
  title = {Real time evolution with neural-network quantum states},
  author = {Guti{\'{e}}rrez, Irene L{\'{o}}pez and Mendl, Christian B.},
  journal = {{Quantum}},
  issn = {2521-327X},
  publisher = {{Verein zur F{\"{o}}rderung des Open Access Publizierens in den Quantenwissenschaften}},
  volume = {6},
  pages = {627},
  month = jan,
  year = {2022}
}

@article{nqstvmc_6,
  doi = {10.22331/q-2023-10-10-1131},
  url = {https://doi.org/10.22331/q-2023-10-10-1131},
  title = {Unbiasing time-dependent {V}ariational {M}onte {C}arlo by projected quantum evolution},
  author = {Sinibaldi, Alessandro and Giuliani, Clemens and Carleo, Giuseppe and Vicentini, Filippo},
  journal = {{Quantum}},
  issn = {2521-327X},
  publisher = {{Verein zur F{\"{o}}rderung des Open Access Publizierens in den Quantenwissenschaften}},
  volume = {7},
  pages = {1131},
  month = oct,
  year = {2023}
}

@article{nqstvmc_7,
  title = {Variational Quantum Dynamics of Two-Dimensional Rotor Models},
  author = {Medvidovi\ifmmode \acute{c}\else \'{c}\fi{}, Matija and Sels, Dries},
  journal = {PRX Quantum},
  volume = {4},
  issue = {4},
  pages = {040302},
  numpages = {16},
  year = {2023},
  month = {Oct},
  publisher = {American Physical Society},
  doi = {10.1103/PRXQuantum.4.040302},
  url = {https://link.aps.org/doi/10.1103/PRXQuantum.4.040302}
}

@Article{nqstvmc_8,
author={Nys, Jannes
and Pescia, Gabriel
and Sinibaldi, Alessandro
and Carleo, Giuseppe},
title={Ab-initio variational wave functions for the time-dependent many-electron Schr{\"o}dinger equation},
journal={Nature Communications},
year={2024},
month={Oct},
day={30},
volume={15},
number={1},
pages={9404},
issn={2041-1723},
doi={10.1038/s41467-024-53672-w},
url={https://doi.org/10.1038/s41467-024-53672-w}
}

@article{nqstvmc_9,
  title={Neural projected quantum dynamics: a systematic study},
  author={Gravina, Luca and Savona, Vincenzo and Vicentini, Filippo},
  journal={Quantum},
  volume={9},
  pages={1803},
  year={2025},
  publisher={Verein zur F{\"o}rderung des Open Access Publizierens in den Quantenwissenschaften}
}

@article{Lange_2024,
doi = {10.1088/2058-9565/ad7168},
url = {https://doi.org/10.1088/2058-9565/ad7168},
year = {2024},
month = {sep},
publisher = {IOP Publishing},
volume = {9},
number = {4},
pages = {040501},
author = {Lange, Hannah and Van de Walle, Anka and Abedinnia, Atiye and Bohrdt, Annabelle},
title = {From architectures to applications: a review of neural quantum states},
journal = {Quantum Science and Technology},
}

@article{liu2021,
  title = {Accurate simulation for finite projected entangled pair states in two dimensions},
  author = {Liu, Wen-Yuan and Huang, Yi-Zhen and Gong, Shou-Shu and Gu, Zheng-Cheng},
  journal = {Phys. Rev. B},
  volume = {103},
  issue = {23},
  pages = {235155},
  numpages = {13},
  year = {2021},
  month = {Jun},
  publisher = {American Physical Society},
  doi = {10.1103/PhysRevB.103.235155},
  url = {https://link.aps.org/doi/10.1103/PhysRevB.103.235155}
}

@article{liu2025hubbard,
  title = {Accurate Simulation of the Hubbard Model with Finite Fermionic Projected Entangled Pair States},
  author = {Liu, Wen-Yuan and Zhai, Huanchen and Peng, Ruojing and Gu, Zheng-Cheng and Chan, Garnet Kin-Lic},
  journal = {Phys. Rev. Lett.},
  volume = {134},
  issue = {25},
  pages = {256502},
  numpages = {8},
  year = {2025},
  month = {Jun},
  publisher = {American Physical Society},
  doi = {10.1103/r4q9-4yvj},
  url = {https://link.aps.org/doi/10.1103/r4q9-4yvj}
}

@article{wu2012phase,
  title = {Phase diagram of the toric code model in a parallel magnetic field},
  author = {Wu, Fengcheng and Deng, Youjin and Prokof'ev, Nikolay},
  journal = {Phys. Rev. B},
  volume = {85},
  issue = {19},
  pages = {195104},
  numpages = {7},
  year = {2012},
  month = {May},
  publisher = {American Physical Society},
  doi = {10.1103/PhysRevB.85.195104},
  url = {https://link.aps.org/doi/10.1103/PhysRevB.85.195104}
}

@article{liu2022gapless,
title = {Gapless quantum spin liquid and global phase diagram of the spin-1/2 ${J}_{1}-{J}_{2}$ square antiferromagnetic \text{Heisenberg} model},
author = {Wen-Yuan Liu and Shou-Shu Gong and Yu-Bin Li and Didier Poilblanc and Wei-Qiang Chen and Zheng-Cheng Gu},
journal = {Science Bulletin},
year = 2022,
volume = {67},
pages = {1034-1041},
issn = {2095-9273},
number = {10},
doi = {https://doi.org/10.1016/j.scib.2022.03.010},
url = {https://www.sciencedirect.com/science/article/pii/S2095927322001001},
}

@article{liu2022emergence,
  title = {Emergence of Gapless Quantum Spin Liquid from Deconfined Quantum Critical Point},
  author = {Liu, Wen-Yuan and Hasik, Juraj and Gong, Shou-Shu and Poilblanc, Didier and Chen, Wei-Qiang and Gu, Zheng-Cheng},
  journal = {Phys. Rev. X},
  volume = {12},
  issue = {3},
  pages = {031039},
  numpages = {17},
  year = {2022},
  month = {Sep},
  publisher = {American Physical Society},
  doi = {10.1103/PhysRevX.12.031039},
  url = {https://link.aps.org/doi/10.1103/PhysRevX.12.031039}
}

@article{liu2024quantum,
  title = {Quantum Criticality with Emergent Symmetry in the Extended Shastry-Sutherland Model},
  author = {Liu, Wen-Yuan and Zhang, Xiao-Tian and Wang, Zhe and Gong, Shou-Shu and Chen, Wei-Qiang and Gu, Zheng-Cheng},
  journal = {Phys. Rev. Lett.},
  volume = {133},
  issue = {2},
  pages = {026502},
  numpages = {9},
  year = {2024},
  month = {Jul},
  publisher = {American Physical Society},
  doi = {10.1103/PhysRevLett.133.026502},
  url = {https://link.aps.org/doi/10.1103/PhysRevLett.133.026502}
}

@article{zhang2025scalable,
  title = {Scalable tensor network algorithm for thermal quantum many-body systems in two dimensions},
  author = {Zhang, Meng and Zhang, Hao and Wang, Chao and He, Lixin},
  journal = {Phys. Rev. B},
  volume = {111},
  issue = {7},
  pages = {075146},
  numpages = {9},
  year = {2025},
  month = {Feb},
  publisher = {American Physical Society},
  doi = {10.1103/PhysRevB.111.075146},
}

@article{wu2025accurate,
  title = {Accurate Gauge-Invariant Tensor-Network Simulations for Abelian Lattice Gauge Theory in $(2+1)\mathrm{D}$: Ground-State and Real-Time Dynamics},
  author = {Wu, Yantao and Liu, Wen-Yuan},
  journal = {Phys. Rev. Lett.},
  volume = {135},
  issue = {13},
  pages = {130401},
  numpages = {9},
  year = {2025},
  month = {Sep},
  publisher = {American Physical Society},
  doi = {10.1103/3m3j-ds18},
  url = {https://link.aps.org/doi/10.1103/3m3j-ds18}
}

@article{cirac2021matrix,
  title = {Matrix product states and projected entangled pair states: Concepts, symmetries, theorems},
  author = {Cirac, J. Ignacio and P\'erez-Garc\'{\i}a, David and Schuch, Norbert and Verstraete, Frank},
  journal = {Rev. Mod. Phys.},
  volume = {93},
  issue = {4},
  pages = {045003},
  numpages = {65},
  year = {2021},
  month = {Dec},
  publisher = {American Physical Society},
  doi = {10.1103/RevModPhys.93.045003},
  url = {https://link.aps.org/doi/10.1103/RevModPhys.93.045003}
}

@article{yang2020observation,
  title={Observation of gauge invariance in a 71-site \text{Bose-Hubbard} quantum simulator},
  author={Yang, Bing and Sun, Hui and Ott, Robert and Wang, Han-Yi and Zache, Torsten V. and Halimeh, Jad C. and Yuan, Zhen-Sheng and Hauke, Philipp and Pan, Jian-Wei},
  journal={Nature},
  volume={587},
  number={7834},
  pages={392--396},
  year={2020},
  url={https://doi.org/10.1038/s41586-020-2910-8},
  publisher={Nature Publishing Group UK London}
}

@Article{Martinez2016,
author={Martinez, Esteban A.
and Muschik, Christine A.
and Schindler, Philipp
and Nigg, Daniel
and Erhard, Alexander
and Heyl, Markus
and Hauke, Philipp
and Dalmonte, Marcello
and Monz, Thomas
and Zoller, Peter
and Blatt, Rainer},
title={Real-time dynamics of lattice gauge theories with a few-qubit quantum computer},
journal={Nature},
year={2016},
month={Jun},
day={01},
volume={534},
number={7608},
pages={516-519},
issn={1476-4687},
doi={10.1038/nature18318},
url={https://doi.org/10.1038/nature18318}
}

@Article{Meth2025,
author={Meth, Michael
and Zhang, Jinglei
and Haase, Jan F.
and Edmunds, Claire
and Postler, Lukas
and Jena, Andrew J.
and Steiner, Alex
and Dellantonio, Luca
and Blatt, Rainer
and Zoller, Peter
and Monz, Thomas
and Schindler, Philipp
and Muschik, Christine
and Ringbauer, Martin},
title={Simulating two-dimensional lattice gauge theories on a qudit quantum computer},
journal={Nature Physics},
year={2025},
month={Apr},
day={01},
volume={21},
number={4},
pages={570-576},
issn={1745-2481},
doi={10.1038/s41567-025-02797-w},
url={https://doi.org/10.1038/s41567-025-02797-w}
}

@misc{SM,
    title = {Supplemental Material},
    author={Wu, Yantao and Nys, Jannes},
    year={2025}
}

@article{hauschild2018efficient,
  title={Efficient numerical simulations with tensor networks: Tensor Network Python (TeNPy)},
  author={Hauschild, Johannes and Pollmann, Frank},
  journal={SciPost Physics Lecture Notes},
  pages={005},
  year={2018}
}

@book{xiang2023tensor, 
  place={Cambridge}, 
  title={Density Matrix and Tensor Network Renormalization}, 
  publisher={Cambridge University Press}, 
  author={Xiang, Tao}, 
  year={2023}}

@article{eisert2010area,
  title = {Colloquium: Area laws for the entanglement entropy},
  author = {Eisert, J. and Cramer, M. and Plenio, M. B.},
  journal = {Rev. Mod. Phys.},
  volume = {82},
  issue = {1},
  pages = {277--306},
  numpages = {0},
  year = {2010},
  month = {Feb},
  publisher = {American Physical Society},
  doi = {10.1103/RevModPhys.82.277},
  url = {https://link.aps.org/doi/10.1103/RevModPhys.82.277}
}

@article{calabrese2016quench,
  title = {Quantum quenches in 1+1 dimensional conformal field theories},
  author = {Calabrese, Pasquale and Cardy, John},
  journal = {Journal of Statistical Mechanics: Theory and Experiment},
  year = {2016},
  month = {jun},
  publisher = {IOP Publishing and SISSA},
  volume = {2016},
  number = {6},
  pages = {064003},
  doi = {10.1088/1742-5468/2016/06/064003},
}

@misc{wu2025algorithm,
      title={Algorithms for variational Monte Carlo calculations of fermion projected entangled pair states in the swap gates formulation and the detailed balance of tensor network sequential sampling},
      author={Yantao Wu and Zhehao Dai},
      year={2025},
      eprint={2506.20106},
      archivePrefix={arXiv},
      primaryClass={cond-mat.str-el},
}

@article{fPEPS_VF,
  title = {Fermionic projected entangled pair states},
  author = {Kraus, Christina V. and Schuch, Norbert and Verstraete, Frank and Cirac, J. Ignacio},
  journal = {Phys. Rev. A},
  volume = {81},
  issue = {5},
  pages = {052338},
  numpages = {6},
  year = {2010},
  month = {May},
  publisher = {American Physical Society},
  doi = {10.1103/PhysRevA.81.052338},
  url = {https://link.aps.org/doi/10.1103/PhysRevA.81.052338}
}

@article{fPEPS_SG,
  title = {Simulation of strongly correlated fermions in two spatial dimensions with fermionic projected entangled-pair states},
  author = {Corboz, Philippe and Or\'us, Rom\'an and Bauer, Bela and Vidal, Guifr\'e},
  journal = {Phys. Rev. B},
  volume = {81},
  issue = {16},
  pages = {165104},
  numpages = {22},
  year = {2010},
  month = {Apr},
  publisher = {American Physical Society},
  doi = {10.1103/PhysRevB.81.165104},
  url = {https://link.aps.org/doi/10.1103/PhysRevB.81.165104}
}

@article{PEPS2004,
  title={Renormalization algorithms for quantum-many body systems in two and higher dimensions},
  author={Verstraete, Frank and Cirac, J. Ignacio},
  journal={arXiv:cond-mat/0407066},
  url={https://doi.org/10.48550/arXiv.cond-mat/0407066},
  year={2004}
}

@article{full_update_1,
  title = {Classical Simulation of Infinite-Size Quantum Lattice Systems in Two Spatial Dimensions},
  author = {Jordan, J. and Or\'us, R. and Vidal, G. and Verstraete, F. and Cirac, J. I.},
  journal = {Phys. Rev. Lett.},
  volume = {101},
  issue = {25},
  pages = {250602},
  numpages = {4},
  year = {2008},
  month = {Dec},
  publisher = {American Physical Society},
  doi = {10.1103/PhysRevLett.101.250602},
  url = {https://link.aps.org/doi/10.1103/PhysRevLett.101.250602}
}

@article{full_update_2,
  title = {Simulation of strongly correlated fermions in two spatial dimensions with fermionic projected entangled-pair states},
  author = {Corboz, Philippe and Or\'us, Rom\'an and Bauer, Bela and Vidal, Guifr\'e},
  journal = {Phys. Rev. B},
  volume = {81},
  issue = {16},
  pages = {165104},
  numpages = {22},
  year = {2010},
  month = {Apr},
  publisher = {American Physical Society},
  doi = {10.1103/PhysRevB.81.165104},
  url = {https://link.aps.org/doi/10.1103/PhysRevB.81.165104}
}

@article{simple_update,
  title = {Accurate Determination of Tensor Network State of Quantum Lattice Models in Two Dimensions},
  author = {Jiang, H. C. and Weng, Z. Y. and Xiang, T.},
  journal = {Phys. Rev. Lett.},
  volume = {101},
  issue = {9},
  pages = {090603},
  numpages = {4},
  year = {2008},
  month = {Aug},
  publisher = {American Physical Society},
  doi = {10.1103/PhysRevLett.101.090603},
  url = {https://link.aps.org/doi/10.1103/PhysRevLett.101.090603}
}

@article{algorithm_finite_peps,
  title = {Algorithms for finite projected entangled pair states},
  author = {Lubasch, Michael and Cirac, J. Ignacio and Ba\~nuls, Mari-Carmen},
  journal = {Phys. Rev. B},
  issue = {6},
  pages = {064425},
  numpages = {16},
  year = {2014},
  month = {Aug},
  publisher = {American Physical Society},
  doi = {10.1103/PhysRevB.90.064425},
  url = {https://link.aps.org/doi/10.1103/PhysRevB.90.064425}
}

@article{TDVP,
  title = {Unifying time evolution and optimization with matrix product states},
  author = {Haegeman, Jutho and Lubich, Christian and Oseledets, Ivan and Vandereycken, Bart and Verstraete, Frank},
  journal = {Phys. Rev. B},
  volume = {94},
  issue = {16},
  pages = {165116},
  numpages = {10},
  year = {2016},
  month = {Oct},
  publisher = {American Physical Society},
  doi = {10.1103/PhysRevB.94.165116},
  url = {https://link.aps.org/doi/10.1103/PhysRevB.94.165116}
}

@article{WII,
  title = {Time-evolving a matrix product state with long-ranged interactions},
  author = {Zaletel, Michael P. and Mong, Roger S. K. and Karrasch, Christoph and Moore, Joel E. and Pollmann, Frank},
  journal = {Phys. Rev. B},
  volume = {91},
  issue = {16},
  pages = {165112},
  numpages = {8},
  year = {2015},
  month = {Apr},
  publisher = {American Physical Society},
  doi = {10.1103/PhysRevB.91.165112},
  url = {https://link.aps.org/doi/10.1103/PhysRevB.91.165112}
}

@article{TEBD,
  title = {Efficient Classical Simulation of Slightly Entangled Quantum Computations},
  author = {Vidal, Guifr\'e},
  journal = {Phys. Rev. Lett.},
  volume = {91},
  issue = {14},
  pages = {147902},
  numpages = {4},
  year = {2003},
  month = {Oct},
  publisher = {American Physical Society},
  doi = {10.1103/PhysRevLett.91.147902},
  url = {https://link.aps.org/doi/10.1103/PhysRevLett.91.147902}
}

@article{iTEBD,
  title = {Classical Simulation of Infinite-Size Quantum Lattice Systems in One Spatial Dimension},
  author = {Vidal, G.},
  journal = {Phys. Rev. Lett.},
  volume = {98},
  issue = {7},
  pages = {070201},
  numpages = {4},
  year = {2007},
  month = {Feb},
  publisher = {American Physical Society},
  doi = {10.1103/PhysRevLett.98.070201},
  url = {https://link.aps.org/doi/10.1103/PhysRevLett.98.070201}
}

@article{zaletel2020isometric,
  title = {Isometric Tensor Network States in Two Dimensions},
  author = {Zaletel, Michael P. and Pollmann, Frank},
  journal = {Phys. Rev. Lett.},
  volume = {124},
  issue = {3},
  pages = {037201},
  numpages = {5},
  year = {2020},
  month = {Jan},
  publisher = {American Physical Society},
  doi = {10.1103/PhysRevLett.124.037201},
  url = {https://link.aps.org/doi/10.1103/PhysRevLett.124.037201}
}

@article{lin2021efficient,
  title = {Efficient simulation of dynamics in two-dimensional quantum spin systems with isometric tensor networks},
  author = {Lin, Sheng-Hsuan and Zaletel, Michael P. and Pollmann, Frank},
  journal = {Phys. Rev. B},
  volume = {106},
  issue = {24},
  pages = {245102},
  numpages = {23},
  year = {2022},
  month = {Dec},
  publisher = {American Physical Society},
  doi = {10.1103/PhysRevB.106.245102},
  url = {https://link.aps.org/doi/10.1103/PhysRevB.106.245102}
}

@article{Fischer2007STMReview,
  title   = {Scanning tunneling spectroscopy of high-temperature superconductors},
  author  = {Fischer, O. and Kugler, M. and Maggio-Aprile, I. and Berthod, C. and Renner, C.},
  journal = {Rev. Mod. Phys.},
  volume  = {79},
  pages   = {353--419},
  year    = {2007},
  doi     = {10.1103/RevModPhys.79.353}
}

@article{Sobota2021ARPESReview,
  title   = {Angle-resolved photoemission spectroscopy},
  author  = {Sobota, Jonathan A. and He, Yu and Shen, Zhi-Xun},
  journal = {Rev. Mod. Phys.},
  volume  = {93},
  number  = {2},
  pages   = {025006},
  year    = {2021},
  doi     = {10.1103/RevModPhys.93.025006}
}

@book{RK,
  title     = {Numerical Methods for Ordinary Differential Equations},
  author    = {Butcher, John C.},
  edition   = {3},
  publisher = {Wiley},
  year      = {2016},
  isbn      = {978-1-119-12499-3},
  doi       = {10.1002/9781119121503}
}

@article{fqhe,
  title = {Two-Dimensional Magnetotransport in the Extreme Quantum Limit},
  author = {Tsui, D. C. and Stormer, H. L. and Gossard, A. C.},
  journal = {Phys. Rev. Lett.},
  volume = {48},
  issue = {22},
  pages = {1559--1562},
  numpages = {0},
  year = {1982},
  month = {May},
  publisher = {American Physical Society},
  doi = {10.1103/PhysRevLett.48.1559},
  url = {https://link.aps.org/doi/10.1103/PhysRevLett.48.1559}
}

@article{laughlin,
  title = {Anomalous Quantum Hall Effect: An Incompressible Quantum Fluid with Fractionally Charged Excitations},
  author = {Laughlin, R. B.},
  journal = {Phys. Rev. Lett.},
  volume = {50},
  issue = {18},
  pages = {1395--1398},
  numpages = {0},
  year = {1983},
  month = {May},
  publisher = {American Physical Society},
  doi = {10.1103/PhysRevLett.50.1395},
  url = {https://link.aps.org/doi/10.1103/PhysRevLett.50.1395}
}

@article{2011fci,
  title   = {Fractional Chern Insulator},
  author  = {Regnault, N. and Bernevig, B. A.},
  journal = {Phys. Rev. X},
  volume  = {1},
  pages   = {021014},
  year    = {2011},
  doi     = {10.1103/PhysRevX.1.021014}
}

@article{thouless1983pump,
  title   = {Quantization of particle transport},
  author  = {Thouless, D. J.},
  journal = {Phys. Rev. B},
  volume  = {27},
  pages   = {6083--6087},
  year    = {1983},
  doi     = {10.1103/PhysRevB.27.6083}
}

@article{fradkin1979phase,
  title   = {Phase diagrams of lattice gauge theories with Higgs fields},
  author  = {Fradkin, Eduardo and Shenker, Stephen H.},
  journal = {Phys. Rev. D},
  volume  = {19},
  pages   = {3682--3697},
  year    = {1979},
  doi     = {10.1103/PhysRevD.19.3682}
}

@article{Sandvik1999boson,
  title   = {Stochastic series expansion method for quantum spin systems},
  author  = {Sandvik, Anders W.},
  journal = {Phys. Rev. B},
  volume  = {59},
  pages   = {R14157--R14160},
  year    = {1999},
  doi     = {10.1103/PhysRevB.59.R14157}
}

@Article{Heinrichs1998,
author={Heinrichs, Stefan
and Mullin, William J.},
title={Quantum-Monte-Carlo Calculations for Bosons in a Two-Dimensional Harmonic Trap},
journal={Journal of Low Temperature Physics},
year={1998},
month={Nov},
day={01},
volume={113},
number={3},
pages={231-236},
issn={1573-7357},
doi={10.1023/A:1022530112706},
url={https://doi.org/10.1023/A:1022530112706}
}

@article{Landau1941English,
  title = {Theory of the Superfluidity of Helium II},
  author = {Landau, L.},
  journal = {Phys. Rev.},
  volume = {60},
  issue = {4},
  pages = {356--358},
  numpages = {0},
  year = {1941},
  month = {Aug},
  publisher = {American Physical Society},
  doi = {10.1103/PhysRev.60.356},
  url = {https://link.aps.org/doi/10.1103/PhysRev.60.356}
}

@article{jain1989composite,
  title = {Composite-fermion approach for the fractional quantum Hall effect},
  author = {Jain, J. K.},
  journal = {Phys. Rev. Lett.},
  volume = {63},
  issue = {2},
  pages = {199--202},
  numpages = {0},
  year = {1989},
  month = {Jul},
  publisher = {American Physical Society},
  doi = {10.1103/PhysRevLett.63.199},
}

@article{sorensen2005fractional,
  title = {Fractional Quantum Hall States of Atoms in Optical Lattices},
  author = {S\o{}rensen, Anders S. and Demler, Eugene and Lukin, Mikhail D.},
  journal = {Phys. Rev. Lett.},
  volume = {94},
  issue = {8},
  pages = {086803},
  numpages = {4},
  year = {2005},
  month = {Mar},
  publisher = {American Physical Society},
  doi = {10.1103/PhysRevLett.94.086803},
}

@article{moller2009composite,
  title = {Composite Fermion Theory for Bosonic Quantum Hall States on Lattices},
  author = {M\"oller, G. and Cooper, N. R.},
  journal = {Phys. Rev. Lett.},
  volume = {103},
  issue = {10},
  pages = {105303},
  numpages = {4},
  year = {2009},
  month = {Sep},
  publisher = {American Physical Society},
  doi = {10.1103/PhysRevLett.103.105303},
}

@article{laughlin1981quantized,
  title = {Quantized Hall conductivity in two dimensions},
  author = {Laughlin, R. B.},
  journal = {Phys. Rev. B},
  volume = {23},
  issue = {10},
  pages = {5632--5633},
  numpages = {0},
  year = {1981},
  month = {May},
  publisher = {American Physical Society},
  doi = {10.1103/PhysRevB.23.5632},
}

@article{czarnik2018time,
  title = {Time evolution of an infinite projected entangled pair state: An algorithm from first principles},
  author = {Czarnik, Piotr and Dziarmaga, Jacek},
  journal = {Phys. Rev. B},
  volume = {98},
  issue = {4},
  pages = {045110},
  numpages = {9},
  year = {2018},
  month = {Jul},
  publisher = {American Physical Society},
  doi = {10.1103/PhysRevB.98.045110},
}

@article{czarnik2019time,
  title = {Time evolution of an infinite projected entangled pair state: An efficient algorithm},
  author = {Czarnik, Piotr and Dziarmaga, Jacek and Corboz, Philippe},
  journal = {Phys. Rev. B},
  volume = {99},
  issue = {3},
  pages = {035115},
  numpages = {12},
  year = {2019},
  month = {Jan},
  publisher = {American Physical Society},
  doi = {10.1103/PhysRevB.99.035115},
}

@Article{hubig2019time,
	title={{Time-dependent study of disordered models with infinite projected entangled pair states}},
	author={Claudius Hubig and J. Ignacio Cirac},
	journal={SciPost Phys.},
	volume={6},
	pages={031},
	year={2019},
	publisher={SciPost},
	doi={10.21468/SciPostPhys.6.3.031},
}

@Article{hubig2020evaluation,
	title={{Evaluation of time-dependent correlators after a local quench in iPEPS: hole motion in the t-J model}},
	author={Claudius Hubig and Annabelle Bohrdt and Michael Knap and Fabian Grusdt and J. Ignacio Cirac},
	journal={SciPost Phys.},
	volume={8},
	pages={021},
	year={2020},
	publisher={SciPost},
	doi={10.21468/SciPostPhys.8.2.021},
}

@article{dziarmaga2021time,
  title = {Time evolution of an infinite projected entangled pair state: Neighborhood tensor update},
  author = {Dziarmaga, Jacek},
  journal = {Phys. Rev. B},
  volume = {104},
  issue = {9},
  pages = {094411},
  numpages = {10},
  year = {2021},
  month = {Sep},
  publisher = {American Physical Society},
  doi = {10.1103/PhysRevB.104.094411},
  url = {https://link.aps.org/doi/10.1103/PhysRevB.104.094411}
}

@article{dziarmaga2022time,
  title = {Time evolution of an infinite projected entangled pair state: A gradient tensor update in the tangent space},
  author = {Dziarmaga, Jacek},
  journal = {Phys. Rev. B},
  volume = {106},
  issue = {1},
  pages = {014304},
  numpages = {8},
  year = {2022},
  month = {Jul},
  publisher = {American Physical Society},
  doi = {10.1103/PhysRevB.106.014304},
  url = {https://link.aps.org/doi/10.1103/PhysRevB.106.014304}
}

@article{ponnaganti2022real,
  title = {Real-time dynamics of a critical resonating valence bond spin liquid},
  author = {Ponnaganti, Ravi Teja and Mambrini, Matthieu and Poilblanc, Didier},
  journal = {Phys. Rev. B},
  volume = {106},
  issue = {19},
  pages = {195132},
  numpages = {15},
  year = {2022},
  month = {Nov},
  publisher = {American Physical Society},
  doi = {10.1103/PhysRevB.106.195132},
  url = {https://link.aps.org/doi/10.1103/PhysRevB.106.195132}
}

@article{evenbly2026loop,
  title = {Loop series expansions for tensor networks},
  author = {Evenbly, Glen and Pancotti, Nicola and Milsted, Ashley and Gray, Johnnie and Chan, Garnet Kin-Lic},
  journal = {Phys. Rev. Res.},
  volume = {8},
  issue = {1},
  pages = {013245},
  numpages = {14},
  year = {2026},
  month = {Mar},
  publisher = {American Physical Society},
  doi = {10.1103/vqks-cr6x},
  url = {https://link.aps.org/doi/10.1103/vqks-cr6x}
}

@Article{tindall2023gauging,
	title={{Gauging tensor networks with belief propagation}},
	author={Joseph Tindall and Matt Fishman},
	journal={SciPost Phys.},
	volume={15},
	pages={222},
	year={2023},
	publisher={SciPost},
	doi={10.21468/SciPostPhys.15.6.222},
	url={https://scipost.org/10.21468/SciPostPhys.15.6.222},
}

@article{espinoza2024spectral,
  title = {Spectral functions with infinite projected entangled-pair states},
  author = {Arias Espinoza, Juan Diego and Corboz, Philippe},
  journal = {Phys. Rev. B},
  volume = {110},
  issue = {9},
  pages = {094314},
  numpages = {8},
  year = {2024},
  month = {Sep},
  publisher = {American Physical Society},
  doi = {10.1103/PhysRevB.110.094314},
  url = {https://link.aps.org/doi/10.1103/PhysRevB.110.094314}
}

@article{motoyama2025tenes,
title = {TeNeS-v2: Enhancement for real-time and finite temperature simulations of quantum many-body systems},
journal = {Computer Physics Communications},
volume = {315},
pages = {109692},
year = {2025},
issn = {0010-4655},
doi = {https://doi.org/10.1016/j.cpc.2025.109692},
url = {https://www.sciencedirect.com/science/article/pii/S0010465525001948},
author = {Yuichi Motoyama and Tsuyoshi Okubo and Kazuyoshi Yoshimi and Satoshi Morita and Tatsumi Aoyama and Takeo Kato and Naoki Kawashima},
}

@article{dai2025fermionic,
  title = {Fermionic Isometric Tensor Network States in Two Dimensions},
  author = {Dai, Zhehao and Wu, Yantao and Wang, Taige and Zaletel, Michael P.},
  journal = {Phys. Rev. Lett.},
  volume = {134},
  issue = {2},
  pages = {026502},
  numpages = {7},
  year = {2025},
  month = {Jan},
  publisher = {American Physical Society},
  doi = {10.1103/PhysRevLett.134.026502},
}

@book{golub2013matrix,
author = {Golub, Gene H. and Van Loan, Charles F.},
title = {Matrix Computations - 4th Edition},
publisher = {Johns Hopkins University Press},
year = {2013},
doi = {10.1137/1.9781421407944},
address = {Philadelphia, PA},
edition   = {},
}

@misc{wu2020dissipative,
      title={Dissipative dynamics in isolated quantum spin chains after a local quench}, 
      author={Yantao Wu},
      year={2020},
      eprint={2010.00700},
      archivePrefix={arXiv},
      primaryClass={cond-mat.stat-mech},
      url={https://arxiv.org/abs/2010.00700}, 
}

@misc{data,
  author       ={Wu, Yantao},
  title        = {github data repository},
  year         = {2025},
  howpublished = {\url{https://github.com/yantaow/open_data/tree/main/wu2025real-time}},
}

@article{sharir2022neural,
  title={Neural tensor contractions and the expressive power of deep neural quantum states},
  author={Sharir, Or and Shashua, Amnon and Carleo, Giuseppe},
  journal={Physical Review B},
  volume={106},
  number={20},
  pages={205136},
  year={2022},
  publisher={APS}
}

@article{hu2025efficient,
  title={Efficient and universal neural-network decoder for stabilizer-based quantum error correction},
  author={Hu, Gengyuan and Ouyang, Wanli and Lu, Chao-Yang and Lin, Chen and Zhong, Han-Sen},
  journal={arXiv preprint arXiv:2502.19971},
  year={2025}
}

@article{sinibaldi2025non,
  title={Non-stabilizerness of Neural Quantum States},
  author={Sinibaldi, Alessandro and Mello, Antonio Francesco and Collura, Mario and Carleo, Giuseppe},
  journal={arXiv preprint arXiv:2502.09725},
  year={2025}
}
\onecolumngrid
\setcounter{equation}{0}
\newpage
\renewcommand{\thesection}{S-\arabic{section}} \renewcommand{\theequation}{S%
\arabic{equation}} \setcounter{equation}{0} \renewcommand{\thefigure}{S%
\arabic{figure}} \setcounter{figure}{0}\setcounter{section}{0}
\centerline{\textbf{Supplemental Material}}
\maketitle
\section{small-$o$ trick}
\footnote{The small-$o$ trick is the following process. 
  Consider $O$ matrix with shape $(N_s, N_p)$. When computing the minSR matrix $OO^\dag$, to save memory, one can partition the parameters into $K$ parts of the same size: i) compute log-derivatives of $1/K$ of the paramters, and form a smaller matrix $o_k$ with shape $(N_s, N_p/K)$ for $k=0,1,2,..,K-1$; ii) form $o_k o_k^\dag$; iii) use an accumulator to obtain the minSR matrix: $OO^\dag = \sum_{k=0}^{K-1} o_k o_k^\dag$.
  $O$ never appears, and one saves the memory by a factor $K$. 
For a generic wavefunction ansatz, e.g. an NQS, computing $o_k$ typically takes more time than $1/K$ of the time needed to compute $O$. 
Then, one needs to spend more time for the blocking.  
For TNS, this does not happen, i.e. there is a structure where one can spend the same amount of time using $o_k$ as using $O$ at once.
In fact, if one is on CPU and does not need to worry about the GPU coding practices, one can also save the time by a factor of $K$.
} 
For tensor network states (TNS), 
for given sample $\vec s$ and at a given lattice site $\vec x$, $O[\vec x](\vec s)_{plrdu} = 0$ if $p \not= \vec s(\vec x)$. 
One thus only keeps 
$$ o[\vec x](\vec s)_{lrdu} = O[\vec x](\vec s)_{\vec s(\vec x)lrdu}$$
This saves memory by a factor of $d$. 
We index $o$ with  $o_{sa}$, where $a$ runs over a selected subset of the parameter sets, depending on the sample $\vec s$.
One can construct the minSR matrix with $o$ instead of $O$, keeping the memory benefit while not sacrificing time cost.
We explain this in detail below.

One needs to form the minSR matrix (also referred to as the neural tangent kernel (NTK)), $T = OO^\dag$. 
To compute this from $o$, one constructs a sample tensor $p$ s.t. $p_{sa} = \vec s(\vec x)$ where $\vec s$ is the sample indexed by $s$ and $\vec x$ is the site associated with the parameter indexed by $a$. 
Then 
\begin{align*}
  \sum_{\alpha} O_{s\alpha} O^\dag_{\alpha s'} &= \sum_{a} o_{sa}o^\dag_{as'} (p_{sa} == p_{s'a})  \\
  &=  \sum_{k=0}^{K-1} \sum_{a} o_{sa}o^\dag_{as'} (p_{sa} == k) \& (p_{s'a} ==  k)  \\
  &= \sum_{k=0}^{K-1} \sum_{a} \tilde{o}(k)_{sa} \tilde{o}(k)^\dag_{as'}
  \intertext{where $\tilde{o}(k)_{sa}  \equiv o_{sa} * (p_{sa}==k)$}
\end{align*}
To implement this, one loops over $k$ and use an accumulator for the sum. 
This saves the memory by a factor of $K$ with no additional time cost. ($oo^\dag$ is $K$-times faster than $OO^\dag$, but one executes $K$ of them in sequence. The memory cost of $p_{sa}$ is negligible because its dtype is \texttt{jnp.int8} compared to $O$'s dtype \texttt{jnp.complex128}).
\begin{lstlisting}[style=mintedpy-doc,caption={JAX implementation of $OO^\dagger$}]
import jax.numpy as jnp

def get_OOdag(o: jnp.ndarray, p: jnp.ndarray, K: int):
    r"""
    Input:
        o : (Ns, Np/K)
        p : (Ns, Np/K)
        K : int
    Return:
        OOdag : (Ns, Ns)
    """
    OOdag = 0
    for k in range(K):
        o2 = jnp.where(p == k, o, 0)
        OOdag += o2 @ o2.T.conj()
    return OOdag
\end{lstlisting} 

\section{Energy evolution}
We use energy conservation as one check for the correctness of our simulations. 
It is worth noting that the TDVP equations fulfil energy conservation by construction, even for ill-estimated preconditioners (see e.g. Ref.~\cite{nqstvmc_3}). 
Hence, predominant sources of violating energy conservation are the finite step size $dt$, the Runge-Kutta integrator, and a limited sample size.

\begin{figure}[t]
  \centering
  \begin{subfigure}[t]{0.45\linewidth}
    \centering
    \includegraphics[width=\linewidth]{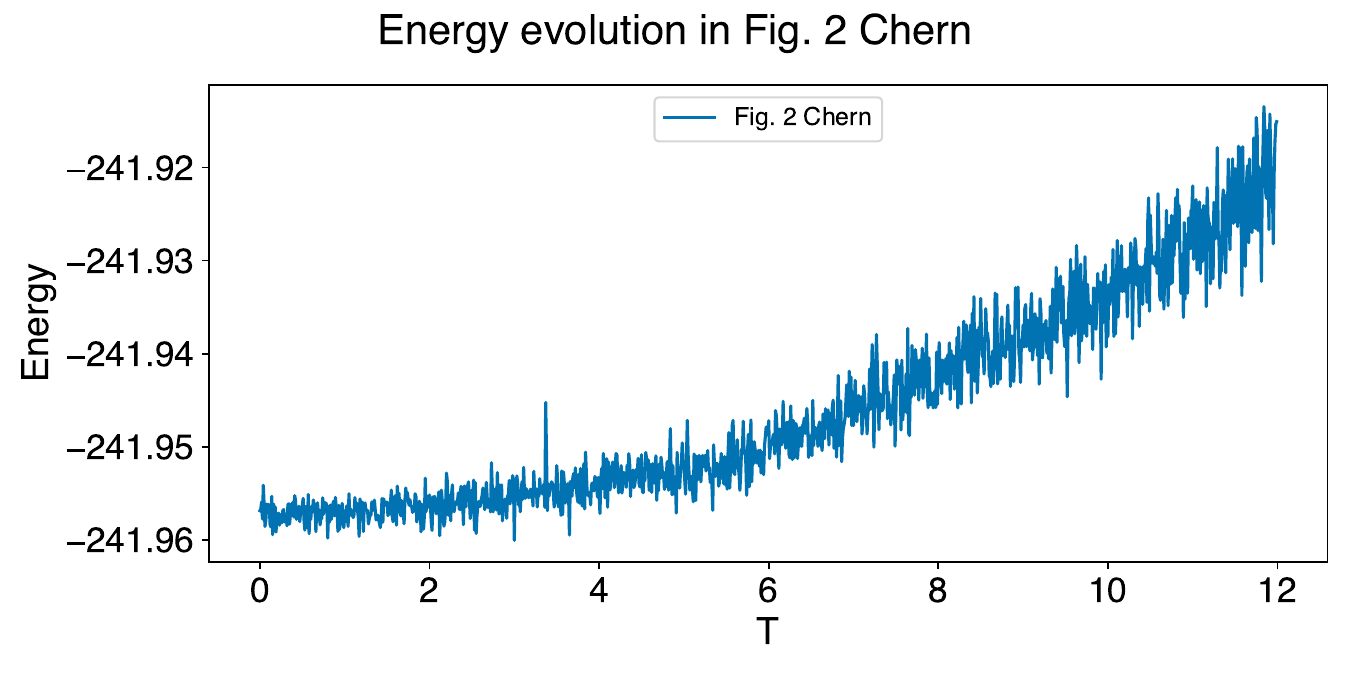}
    \caption{Energy evolution of the Chern insulator.
The relative energy drift is $2\times 10^{-4}$ over the evolution.
    }
    \label{fig:e_chern}
  \end{subfigure}\hfill
  \begin{subfigure}[t]{0.45\linewidth}
    \centering
    \includegraphics[width=\linewidth]{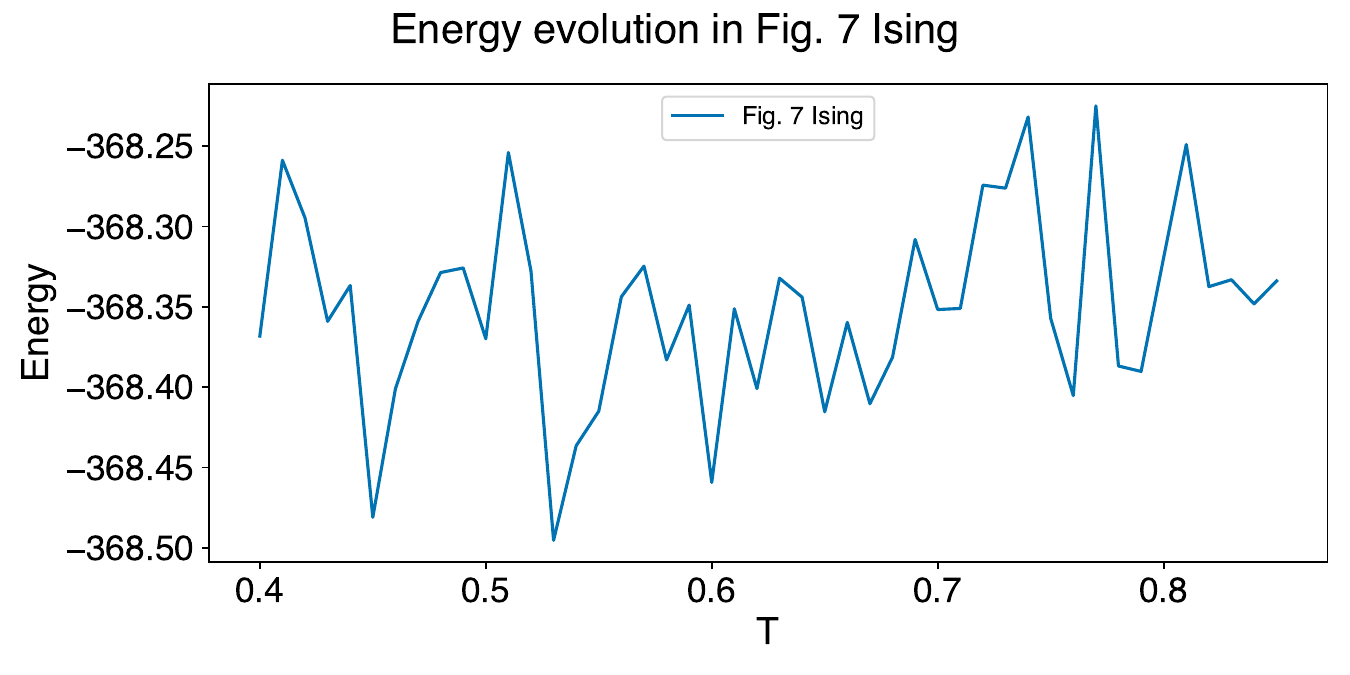}
    \caption{Energy evolution of the Ising model during global quench.
    The initial energy can be computed exactly: $-L^2h = -11^2 \times 3.04438 = -368.36998$.}
    \label{fig:e_ising}
  \end{subfigure}
  \begin{subfigure}[t]{0.45\linewidth}
    \centering
    \includegraphics[width=\linewidth]{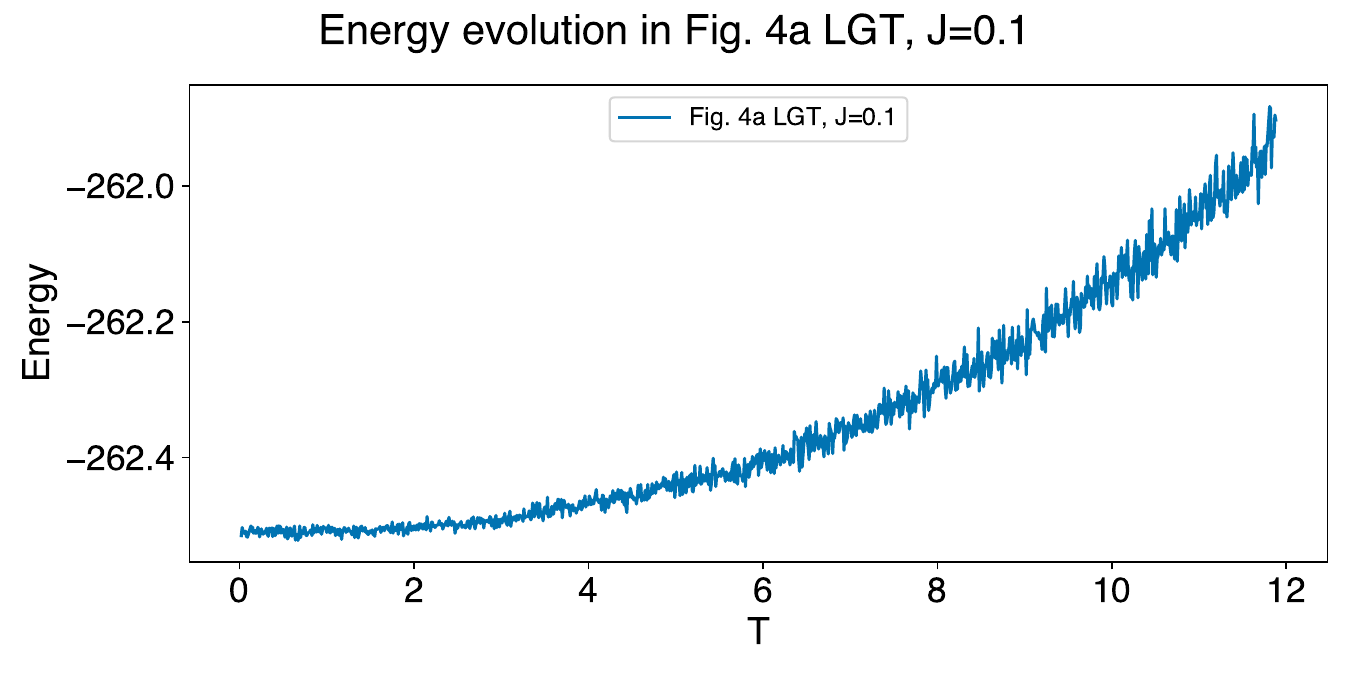}
    \caption{Energy evolution of the $\Z_2$ lattice gauge theory in the deconfined phase. 
The relative energy drift is $1\times 10^{-3}$ over the evolution.}
    \label{fig:e_LGT_deconfined1}
  \end{subfigure}\hfill
  \begin{subfigure}[t]{0.45\linewidth}
    \centering
    \includegraphics[width=\linewidth]{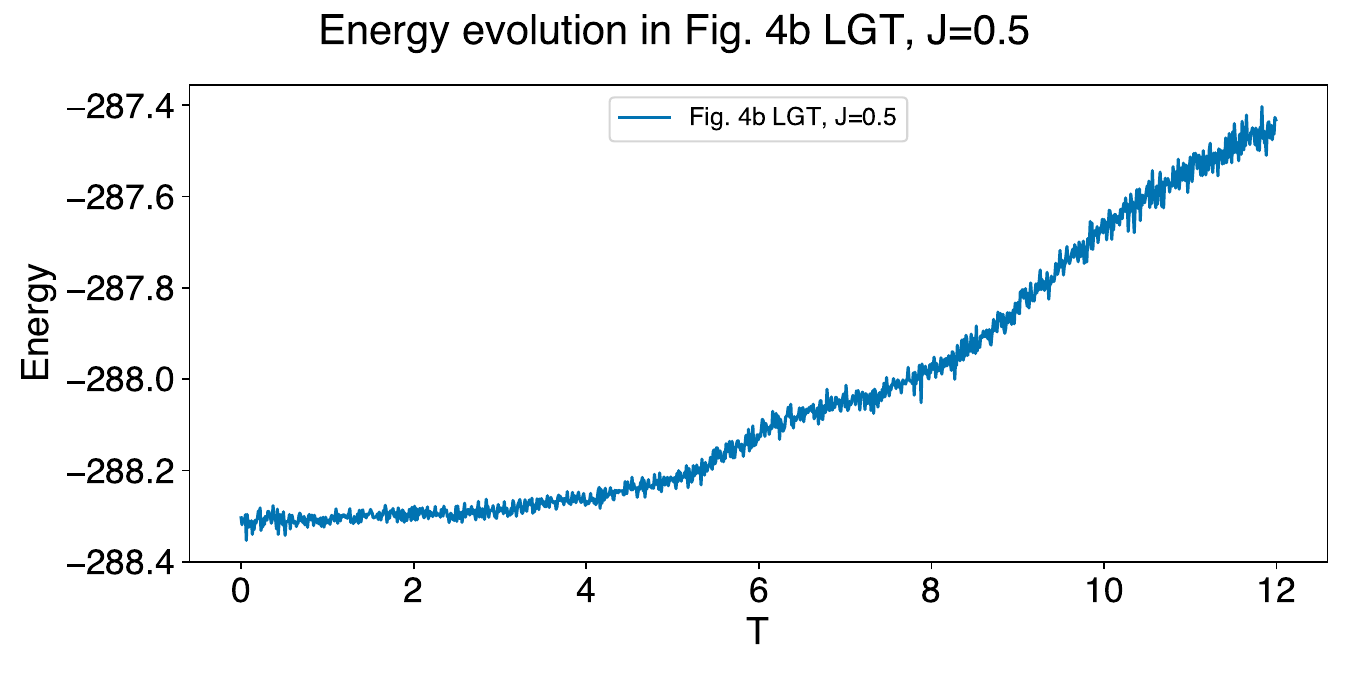}
    \caption{Energy evolution of the $\Z_2$ lattice gauge theory in the Higgs phase. 
The relative energy drift is $2\times 10^{-3}$ over the evolution.}
    \label{fig:e_LGT_deconfined2}
  \end{subfigure}
  \caption{Energy evolutions of the simulations presented in the main text. }
\end{figure}

\end{document}